\documentclass[final,1p,times,authoryear]{elsarticle}

\usepackage{booktabs}

\makeatletter
\def\ps@pprintTitle{%
 \let\@oddhead\@empty
 \let\@evenhead\@empty
 \def\@oddfoot{}%
 \let\@evenfoot\@oddfoot}
\makeatother

\usepackage{graphicx}
\usepackage{amssymb}
\usepackage{commath}
\usepackage{mathtools}
\usepackage{subcaption}
\usepackage{float}
\usepackage{natbib}
\usepackage{hyperref}

\graphicspath{ {./graphics/} }

\begin{document}

\begin{frontmatter}
\title{Bayesian Modeling and Estimation of Linear Time-Varying Systems using Neural Networks and Gaussian Processes}
\author{Yaniv Shulman}
\ead{yaniv@shulman.info}

\begin{abstract}
The identification of Linear Time-Varying (LTV) systems from input-output data is a fundamental yet challenging ill-posed inverse problem. This work introduces a unified Bayesian framework that models the system's impulse response, $h(t, \tau)$, as a stochastic process. We decompose the response into a posterior mean and a random fluctuation term, a formulation that provides a principled approach for quantifying uncertainty, unifies intrinsic channel variability and epistemic uncertainty through a common posterior representation, and naturally defines a new, useful system class we term Linear Time-Invariant in Expectation (LTIE). To perform inference, we leverage modern machine learning techniques, including Bayesian neural networks and Gaussian Processes, using scalable variational inference. We demonstrate through a series of experiments that our framework can infer the properties of an LTI system from a single noisy input-output pair, including under deliberate additive-noise misspecification, achieve a lower overall error floor than the classical CCF stacking baseline in a simulated ambient noise tomography setting, and track a continuously varying LTV impulse response by using a structured Gaussian Process prior. This work provides a flexible and robust methodology for uncertainty-aware system identification in dynamic environments.
\end{abstract}

\end{frontmatter}

\section{Introduction}
\label{s:Introduction}

Linear Time-Varying (LTV) systems are fundamental to modeling dynamic processes in fields ranging from geophysics and communications to control theory \citep{Kozachek2024, Lin2020, Bensen2007}. Unlike their time-invariant counterparts, an LTV system's behavior is described by an impulse response, $h(t, \tau)$, that changes over time, posing significant challenges for analysis and estimation \citep{Kailath1962, Bello1963}. The task of identifying $h(t, \tau)$ from input-output data is a severely ill-posed inverse problem, as one must infer a function of two variables from one-dimensional time series \citep{Aubel2015}. This work introduces a Bayesian framework for modeling such systems, where the inherent uncertainty and time-varying nature are captured probabilistically.

Our central approach is to treat the impulse response itself as a stochastic process. We reparameterize the LTV impulse response into two components: a deterministic posterior mean, which represents the system's average behavior conditioned on observations, and a zero-mean stochastic process, which captures the random fluctuations or innovations. This decomposition, $h = \mu + \mathcal{E}$, provides a powerful and generalizable modeling framework. It allows us to unify the description of a wide spectrum of system behaviors, from purely deterministic systems to fully random ones, under a single probabilistic lens. The decomposition itself is not mathematically novel once a law over $h$ is specified; rather, the contribution here is to adopt it explicitly as a posterior representation for LTV system identification. More concretely, the contributions of this paper are: a posterior interpretation of stochastic impulse-response models for system identification; the LTIE-law viewpoint for inference over deterministic but unknown LTI systems; practical variational estimators for LTI and LTV impulse-response regression; and synthetic demonstrations of uncertainty-aware estimation in single-observation deconvolution, ambient noise tomography, and continuously varying LTV settings.

Within this framework, we define a useful class of laws over impulse responses, which we term \emph{Linear Time-Invariant in Expectation} (LTIE).  In an LTIE channel model, the mean impulse response is constant, $\mu(t, \tau) = \mu(\tau)$, but the system still exhibits stochastic fluctuations. This concept is related to, but more general than, the "mean LTI" or "line-of-sight" component in channel models like Rician fading, as our framework allows for arbitrary, non-stationary covariance structures for the fluctuations around this constant mean. The LTIE concept provides a crucial conceptual bridge: when inference is performed in a model class that enforces or symmetrically favors time-invariant impulse responses, a deterministic but unknown LTI system induces an LTIE posterior law over impulse responses whose mean is time-invariant. In this sense, intrinsic channel variability and epistemic uncertainty are unified through the same posterior representation $h=\mu+\mathcal{E}$, even though they correspond to different underlying sources of randomness before conditioning on data. The novelty of this work lies not in positing a constant-mean random channel per se, but in the Bayesian posterior interpretation of this representation and in leveraging the generalized framework with modern machine learning tools, namely Bayesian neural networks and Gaussian processes, to perform robust, uncertainty-aware estimation of the impulse response. In the following sections, we develop this framework for both continuous and discrete-time systems, deriving key statistical properties and discussing their implications.

\section{Related Work}

The conceptual approach of this work, which models the impulse response as a stochastic process, builds upon the seminal statistical characterization of randomly time-varying linear channels by Bello \citep{Bello1963}. This foundational research led to widely used models such as the Wide-Sense Stationary Uncorrelated Scattering (WSSUS) channel. In our decomposition $h=\mu+\mathcal{E}$, WSSUS is most naturally described by a time-invariant mean, $\mu(t,\tau)=\mu(\tau)$, together with wide-sense stationarity and uncorrelated scattering assumptions on the centered fluctuation process $\mathcal{E}$; the common zero-mean special case is recovered by setting $\mu=0$ \citep{Yoo2005}. The Rician fading model is another well-known special case that includes a deterministic, line-of-sight component alongside random, scattered components, which is conceptually analogous to our decomposition of the impulse response $h$ into a non-zero mean component $\mu$ and a stochastic fluctuation $\mathcal{E}$ \citep{SimonAlouini2005}. Our framework provides a generalization of these classic models by accommodating a time-varying mean and arbitrary, non-stationary covariance structures.

Modern Bayesian system identification has moved towards placing priors directly on the impulse response itself. A significant development in this area is kernel-based regularization, where the impulse response is modeled as a draw from a Gaussian Process (GP) \citep{Pillonetto2023, Darwish2017}. This function-space perspective allows prior knowledge to be encoded via the GP's covariance function, or kernel. For instance, specialized "stable kernels" have been designed to enforce physical properties like smooth exponential decay, providing a principled form of regularization that ensures the stability of the identified model \citep{Pillonetto2023}.

The flexibility of machine learning has also been brought to bear on this problem. Neural networks (NNs) are widely used as universal function approximators for nonlinear systems, often in autoregressive configurations like NARX \citep{Narendra1990, Nelles2020}. To address the fact that standard NNs provide only point estimates, Bayesian Neural Networks (BNNs) have been developed, which place prior distributions on network weights to infer a full posterior distribution over models, thereby capturing model uncertainty \citep{Jospin2022, Zhou2022}. Physics-Informed BNNs (BPINNs) extend this by incorporating governing physical equations into the loss function, a technique that has proven robust for system identification in the presence of noise \citep{Stock2024}.

The synergy between GPs and NNs is an active area of research. While some work focuses on constructing deep hierarchies of GP mappings (Deep GPs) \citep{Damianou2013} or designing BNNs that replicate GP priors \citep{Sendera2025}, other approaches use GPs as structured priors within a larger Bayesian model to capture temporal dependencies. For example, recent work has combined GP priors on system states with dynamics governed by Neural Ordinary Differential Equations (ODEs) \citep{Bhouri2022}, and GP regression has also been used directly for nonparametric identification of linear time-varying systems \citep{Hallemans2020}. This work adopts a similar philosophy, using a GP as a structured prior within a neural amortized variational inference model whose deterministic encoder parameterizes a Bayesian latent impulse-response layer. The practical application of these advanced Bayesian models is made possible by approximate inference techniques, as the true posterior distribution is almost always intractable. Variational Inference (VI) has emerged as a computationally efficient and scalable alternative to sampling-based methods, and has proven highly effective for deconvolution problems, which are mathematically equivalent to the impulse response estimation task addressed in this paper \citep{Zhang2013, Susik2023, Jospin2022}.

It is important to distinguish the present framework from the standard Bayesian observation model used in LTI system identification. In that setting, one typically writes the observations as a deterministic convolution corrupted by additive noise, so that uncertainty enters through both a posterior distribution over the impulse response and an explicit observation-noise model \citep{Pillonetto2023, Darwish2017}. By contrast, in our framework the primary latent random object is the impulse response itself. This yields a different interpretation of predictive uncertainty: in the standard additive-noise model, the output variance typically contains both a signal-dependent term due to posterior uncertainty in the impulse response and a signal-independent observation-noise floor, whereas in our stochastic-channel model the induced output variance is entirely channel-driven and therefore depends on the input signal that probes the system. This distinction is central to the interpretation of LTIE posterior laws: when inference is performed in a model class that enforces or symmetrically favors time-invariant impulse responses, a deterministic but unknown LTI system induces a posterior law whose mean is time-invariant even though the inferred output uncertainty remains signal-dependent.

In the following sections, we develop this framework for both continuous and discrete-time systems, validate its performance through a series of increasingly complex estimation tasks, and conclude with a discussion of its implications and avenues for future work.

\section{A Bayesian Framework for LTV Systems}
\label{s:bayesian_framework}

We formalize the analysis of LTV systems by treating the impulse response as a stochastic process within a Bayesian setting. The system is probed by a known, deterministic input signal. Our primary modeling choice is to place the uncertainty in the impulse response itself, so that randomness is represented as an intrinsic property of the system or medium. When needed, an additional observation model can be layered on top of this latent channel model to account for measurement noise. This distinction allows the same posterior representation to describe both genuinely random channels and uncertainty about a deterministic but unknown channel. A key consequence is that the uncertainty induced at the output is shaped by the probing input through the posterior covariance of the impulse response, rather than arising solely as an additive signal-independent noise floor.

\subsection{Continuous-Time Systems}

A continuous LTV system maps a deterministic input signal $s(t)$ to a latent channel output $r(t)$ via the convolution integral:
\begin{equation}
r(t) = \int_{-\infty}^{\infty} h(t, \tau) \, s(t - \tau) \, d\tau,
\end{equation}
where $h(t, \tau)$ is the system's response at time $t$ to an impulse applied at time $t-\tau$. For causal systems, $h(t, \tau) = 0$ for $\tau < 0$. When measurement noise is present, the observed signal is modeled as
\begin{equation}
y(t) = r(t) + \nu(t),
\end{equation}
where $\nu(t)$ is an observation-noise process. Unless stated otherwise, the analysis below concerns the latent channel output $r(t)$ and the special case $\nu(t)\equiv 0$. Signals and impulse responses may be real- or complex-valued throughout; in the real-valued case, conjugation symbols may be omitted.

In our Bayesian model, the impulse response is a stochastic process. Conditioned on a set of observed data $\mathcal{D}$, we decompose it as:
\begin{equation}
h(t, \tau) = \mu(t, \tau) + \mathcal{E}(t, \tau),
\end{equation}
where $\mu(t, \tau) = \mathbb{E}[h(t, \tau) \mid \mathcal{D}]$ is the \emph{posterior mean} impulse response, capturing the system's expected behavior. The term $\mathcal{E}(t, \tau)$ is a zero-mean stochastic process, $\mathbb{E}[\mathcal{E}(t, \tau) \mid \mathcal{D}] = 0$, representing the \emph{posterior fluctuations} or random deviations from that mean. By construction, the posterior mean and fluctuation components are uncorrelated.

To ensure that the convolution and second-order moments are well defined, we assume throughout that, for each fixed $t$, the function $\tau \mapsto h(t,\tau)$ is integrable (or square-integrable, as appropriate for the signals under consideration), that the deterministic input $s$ is chosen so that the convolution integral exists, and that the fluctuation process $\mathcal{E}(t,\tau)$ has finite second moments. We further assume the standard measurability and integrability conditions required for exchanging expectation and integration (e.g., by Fubini's or Tonelli's theorem). Under these assumptions, the expected output of the system is governed solely by the posterior mean of the impulse response. By linearity of the expectation operator, we have:
\begin{equation}
\mathbb{E}\left[ r(t) \mid \mathcal{D} \right] = \int_{-\infty}^{\infty} \mathbb{E}[h(t, \tau) \mid \mathcal{D}] \, s(t - \tau) \, d\tau = \int_{-\infty}^{\infty} \mu(t, \tau) \, s(t - \tau) \, d\tau.
\end{equation}
On average, the system behaves like a deterministic LTV system with impulse response $\mu(t, \tau)$.

The variability of the output signal arises from the stochastic fluctuations $\mathcal{E}(t, \tau)$. The variance of $r(t)$ is given by:
\[
\operatorname{Var}[r(t) \mid \mathcal{D}] = \mathbb{E}\left[ \left| r(t) - \mathbb{E}[r(t) \mid \mathcal{D}] \right|^2 \mid \mathcal{D} \right] = \mathbb{E}\left[\left|\int_{-\infty}^{\infty} \mathcal{E}(t, \tau) \, s(t-\tau) \, d\tau\right|^2 \mid \mathcal{D}\right].
\]
Assuming the necessary conditions for swapping expectation and integration (via Fubini's theorem), this variance can be expressed as a function of the input signal and the posterior covariance of the fluctuations:
\[
\operatorname{Var}[r(t) \mid \mathcal{D}] = \int_{-\infty}^{\infty} \int_{-\infty}^{\infty} R_{\mathcal{E}}(t; \tau, \tau') \, s(t-\tau) s^*(t-\tau') \, d\tau \, d\tau',
\]
where
\[
R_{\mathcal{E}}(t; \tau, \tau') = \mathbb{E}\!\left[\mathcal{E}(t,\tau)\mathcal{E}^*(t,\tau') \mid \mathcal{D}\right]
\]
is the posterior covariance kernel of the zero-mean fluctuation process $\mathcal{E}(t,\tau)$ at fixed time $t$. Since $\mathbb{E}[\mathcal{E}(t,\tau)\mid\mathcal{D}] = 0$, this is equivalently the conditional covariance kernel. This expression quantifies how uncertainty in the impulse response translates to uncertainty in the output.

If an additive observation-noise term $\nu(t)$ is included and is conditionally zero-mean and conditionally uncorrelated with the impulse-response fluctuations, then the observed signal satisfies
\[
\mathbb{E}[y(t)\mid\mathcal{D}] = \mathbb{E}[r(t)\mid\mathcal{D}],
\qquad
\operatorname{Var}[y(t)\mid\mathcal{D}] = \operatorname{Var}[r(t)\mid\mathcal{D}] + \operatorname{Var}[\nu(t)\mid\mathcal{D}].
\]
Thus the stochastic impulse-response model is the latent system model, while additive measurement noise can be incorporated as a separate observation layer when appropriate.

The system's behavior can also be analyzed in the frequency domain. If the posterior mean is time-invariant, $\mu(t,\tau)=\mu(\tau)$, the system is \emph{Linear Time-Invariant in Expectation} (LTIE). For deterministic probing, a natural spectral characterization is obtained from repeated realizations of the random medium. Let $r_m^{(T)}(t)$ denote the output observed on the interval $[0,T]$ for the $m$-th realization, and define its finite-window Fourier transform

\begin{equation}
R_m^{(T)}(f)=\int_0^T r_m^{(T)}(t)e^{-j2\pi ft}\,dt.
\end{equation}

The averaged periodogram
\begin{equation}
\widehat{S}_M^{(T)}(f)=\frac{1}{MT}\sum_{m=1}^M \abs{R_m^{(T)}(f)}^2
\end{equation}
converges, for each fixed observation window $T$ and each fixed frequency $f$, under standard independence assumptions across realizations (or, more generally, under an appropriate law of large numbers for dependent trials) and provided $\mathbb{E}[\abs{R_1^{(T)}(f)}^2 \mid \mathcal{D}]<\infty$, to the finite-time ensemble spectrum
\begin{equation}
S_r^{(T)}(f)=\frac{1}{T}\mathbb{E}[\abs{R_1^{(T)}(f)}^2 \mid \mathcal{D}].
\end{equation}

This provides a natural finite-time ensemble spectral description for deterministic probing through an LTIE channel. When the randomness represents intrinsic channel variability, the realizations $\{r_m^{(T)}\}_{m=1}^M$ are physical repeated trials; when the randomness instead represents posterior uncertainty about an otherwise deterministic system, these realizations are more appropriately interpreted as outputs induced by posterior draws of the impulse response under the fixed probe. In the latter case, $S_r^{(T)}(f)$ is best viewed as a posterior predictive second spectral moment rather than, necessarily, as a physical ensemble PSD. If, in addition, the centered output process admits a wide-sense stationary infinite-time limit, then this finite-time spectrum reduces to the ordinary PSD. In the more general LTV case where $\mu(t,\tau)$ depends on $t$, the system exhibits non-stationary behavior in its mean, and time-frequency analysis methods are required to characterize the output spectrum. A proof is omitted, as the convergence follows directly from the law of large numbers applied pointwise in frequency to $\abs{R_m^{(T)}(f)}^2$.

\subsection{Discrete-Time Systems}

The framework extends naturally to discrete-time systems, which are often described by a Finite Impulse Response (FIR) model. For notational convenience, the discrete-time theory and experiments below use $(f,g,y)$ for input, latent output, and observed output respectively, in place of the continuous-time symbols $(s,r,y)$. 

The latent channel output $g[n]$ in response to a deterministic input sequence $f[n]$ is given by the convolution sum:
\begin{equation}
g[n] = \sum_{k=1}^{p} h_k[n] f[n - k].
\end{equation}
When measurement noise is present, the observed signal is modeled as
\begin{equation}
y[n] = g[n] + \nu[n],
\end{equation}
where $\nu[n]$ is an observation-noise process. Here, the impulse response at time $n$ is a $p$-dimensional random vector $h[n]=(h_1[n], \ldots, h_p[n])^\top$, where $h_k[n]$ denotes the $k$-th tap at time $n$; equivalently, in elementwise notation we write the time-varying impulse response as $h[n,k] \equiv h_k[n]$. The summation begins at $k=1$, following a common convention in system identification where the direct feedthrough term ($k=0$) is often handled separately or assumed to be zero. Unless stated otherwise, the analysis below concerns the latent channel output $g[n]$ and the special case $\nu[n]\equiv 0$.

Following our Bayesian approach, we decompose the impulse response vector as:
\begin{equation}
h[n] = \mu[n] + \mathcal{E}[n],
\end{equation}
where $\mu[n] = \mathbb{E}[h[n]\mid\mathcal{D}]$ is the posterior mean vector and $\mathcal{E}[n]$ is the zero-mean vector of posterior fluctuations. The received signal can then be written as the sum of a mean component and a stochastic component:
\begin{equation}
g[n] = \underbrace{\sum_{k=1}^{p} \mu_k[n] f[n - k]}_{g_{\text{mean}}[n]} + \underbrace{\sum_{k=1}^{p} \mathcal{E}_k[n] f[n - k]}_{g_{\mathcal{E}}[n]}.
\end{equation}
The expected value of the received signal is simply its mean component, $\mathbb{E}[ g[n] \mid \mathcal{D} ] = g_{\text{mean}}[n]$.

If an additive observation-noise term $\nu[n]$ is included and is conditionally zero-mean and conditionally uncorrelated with the impulse-response fluctuations, then
\[
\mathbb{E}[y[n]\mid\mathcal{D}] = \mathbb{E}[g[n]\mid\mathcal{D}],
\qquad
\operatorname{Var}[y[n]\mid\mathcal{D}] = \operatorname{Var}[g[n]\mid\mathcal{D}] + \operatorname{Var}[\nu[n]\mid\mathcal{D}].
\]

The variance of the output at time $n$ depends on the conditional second-moment matrix (equivalently, the conditional covariance matrix, since $\mathcal{E}[n]$ is zero-mean) of the impulse response fluctuations at that same time. Let $\mathbf{f}[n]=(f[n-1], \ldots, f[n-p])^\top$ be the vector of recent inputs, and let $\Sigma[n]=\operatorname{Cov}[h[n]\mid\mathcal{D}] = \mathbb{E}[\mathcal{E}[n]\mathcal{E}[n]^{H} \mid \mathcal{D}]$ be the $p \times p$ posterior covariance matrix of the impulse response, where $(\cdot)^H$ denotes the Hermitian transpose. The output variance is then:
\begin{equation}
\operatorname{Var}[ g[n] \mid \mathcal{D} ] = \operatorname{Var}[ \mathbf{f}[n]^\top \mathcal{E}[n] \mid \mathcal{D} ] = \mathbf{f}[n]^\top \Sigma[n] \mathbf{f}[n]^*.
\end{equation}
This shows explicitly that, in the latent stochastic-channel model, the output uncertainty is signal-dependent and vanishes in the absence of excitation.

Similarly, the covariance of the received signal across different times $n$ and $m$ depends on the cross-time conditional second moment of the impulse-response fluctuations. Define
\[
R_{\mathcal{E}}[n,m] = \mathbb{E}\!\left[\mathcal{E}[n]\mathcal{E}[m]^H \mid \mathcal{D}\right],
\]
whose $(k,l)$ entry is
\[
R_{\mathcal{E},kl}[n,m] = \mathbb{E}\!\left[\mathcal{E}_k[n]\mathcal{E}_l[m]^* \mid \mathcal{D}\right].
\]
Since $\mathcal{E}[n]$ is conditionally zero-mean, this is equivalently the cross-time conditional covariance matrix. The output covariance is therefore
\begin{equation}
\operatorname{Cov}[ g[n], g[m] \mid \mathcal{D} ] = \sum_{k=1}^{p} \sum_{l=1}^{p} f[n - k] f^*[m - l] \, R_{\mathcal{E},kl}[n,m].
\end{equation}
If the fluctuations are uncorrelated across time, i.e., $R_{\mathcal{E}}[n,m] = \delta_{n,m} \Sigma[n]$, the output signal $g[n]$ will have a non-stationary white stochastic component.

Under a local stationarity assumption, where the statistics of $h[n]$ vary slowly enough to be considered approximately constant over a short-time analysis window centered at time $n$, and with boundary effects within that window taken to be negligible, we can analyze the system using a local time-frequency description \citep{Priestley1965, Dahlhaus1997}. Let $\widetilde{F}(n,e^{j\omega})$ denote a windowed transform of the deterministic input over that local window, let
 \[
 \mu_n(e^{j\omega}) \coloneqq \sum_{k=1}^{p} \mu_k[n] e^{-j\omega k}
 \]
denote the DTFT of the local mean impulse response at time $n$, and define the local random fluctuation transfer function
\[
H_{\mathcal{E},n}(e^{j\omega}) \coloneqq \sum_{k=1}^{p} \mathcal{E}_k[n] e^{-j\omega k},
\]
with local second spectral moment
\[
S_{H_{\mathcal{E}}}(n,e^{j\omega}) \coloneqq \mathbb{E}\!\left[\abs{H_{\mathcal{E},n}(e^{j\omega})}^2 \mid \mathcal{D}\right].
\]
Then a heuristic short-time approximation to the local output spectrum is

\begin{equation}
\widetilde{S}_{gg}(n, e^{j\omega}) \approx \abs{\widetilde{F}(n,e^{j\omega})}^2 \left( \abs{\mu_n(e^{j\omega})}^2 + S_{H_{\mathcal{E}}}(n, e^{j\omega}) \right).
\end{equation}
Here, $\widetilde{S}_{gg}(n, e^{j\omega})$ denotes a local finite-window spectral characterization. This approximation is not claimed as an exact identity outside the local stationarity regime; it relies on the zero-mean property of the fluctuations to remove the cross term in the local second moment and on the window being short enough that within-window non-stationarity is negligible.

\subsection{Summary}

This section presents a Bayesian framework for modeling Linear Time-Varying systems by reparameterizing the impulse response into a posterior mean and a stochastic fluctuation term, $h = \mu + \mathcal{E}$. This decomposition provides a conceptually clear and mathematically tractable method for analyzing systems with inherent uncertainty and non-stationarity. A key strength of the framework is that the same posterior representation can be interpreted in two complementary ways: as a model for genuinely random channels, or as a representation of epistemic uncertainty about a deterministic but unknown channel. In the latter case, if the posterior mean is time-invariant, the inferred posterior law is naturally LTIE. Additive observation noise, when relevant, can be incorporated as a separate observation layer on top of this latent system model.

The framework provides a unified perspective that connects deterministic LTV systems, LTI systems, and fully stochastic LTV systems. The degree of time variation in the posterior mean $\mu(t, \tau)$ and the structure of the posterior covariance $R_{\mathcal{E}}$ determine where a specific system lies on this spectrum. This flexible approach provides a robust foundation for the regression and estimation of impulse responses, offering a principled way to quantify uncertainty and predict system behavior in dynamic environments. It is important to note that, as with any system identification problem, the ability of the data to identify the impulse response $h$ and concentrate the posterior distribution $p(h \mid \mathcal{D})$ depends on the properties of the input signal and on the chosen model class. Qualitatively, the input must be persistently exciting across the modes of interest in the system, ensuring that all aspects of the impulse response are sufficiently probed \citep{Willems2005PE}. In the practical estimators used in our experiments, identifiability is aided not only by excitation but also by structural restrictions such as finite impulse-response support (or kernel-size truncation) and prior regularization, which reduce the effective hypothesis space. While a detailed analysis of identifiability is beyond the scope of this work, our experimental results demonstrate that for sufficiently rich input signals and appropriate structural regularization, the posterior distribution is well-constrained and provides a meaningful estimate of the system and its uncertainty. Throughout, the phrase ``posterior fluctuations'' refers to randomness in the inferred law of $h$ after conditioning on data; depending on context, this may represent either intrinsic channel variability or epistemic uncertainty about an otherwise deterministic system.

\section{Example Applications}

The Bayesian framework developed in this paper is applicable to any field where characterizing the transformation between an input and an output signal is critical. By modeling the system probabilistically, we can improve signal quality, predict outcomes with quantified uncertainty, and infer latent system properties from observed data. The source code, data, demonstrative Jupyter notebooks, and environment specifications for the following experiments are available online at \url{https://github.com/yaniv-shulman/bayes-ltv}. Unless stated otherwise, the experiments in this section should be interpreted primarily as validations of the usefulness of the posterior representation and associated estimators, rather than as direct empirical confirmation that the latent stochastic impulse-response model is the literal physical data-generating mechanism. In particular, when data are generated under a deterministic impulse response with additive observation noise, the resulting posterior law over impulse responses is best viewed as an uncertainty-aware inferential representation under deliberate model misspecification. Unless stated otherwise, terminology such as ``posterior mean,'' ``posterior variance,'' ``posterior samples,'' and ``posterior predictive'' in the experimental sections refers to the variational or otherwise approximate posterior returned by the estimator used in that experiment, not to an analytically exact posterior. For readability, we often use the shorter term ``posterior'' once this distinction has been made. Key experiment-defining settings, including noise levels, optimizer schedules, GP hyperparameters, windowing choices, and the definitions of the plotted uncertainty bands, are summarized in~\ref{app:experimental_settings}.

\subsection{Impulse Response Regression from a Single Observation}

As a foundational example, we demonstrate how the Bayesian machinery can robustly solve the classic problem of system identification from a single, noisy observation pair. While the theoretical sections analyzed a latent stochastic impulse-response model, this experiment addresses a common precursor: how to infer the properties of a deterministic LTI system when they are unknown and masked by additive observation noise. This experiment is intentionally included as a misspecification test: the data are generated under the more conventional additive-noise model rather than under a stochastic impulse-response model, in order to show that the same Bayesian framework remains useful and robust even when its latent noise interpretation is not the literal data-generating mechanism.

The goal is to move beyond a single point estimate of the impulse response and instead recover a variational approximation, $q(h \mid f, y)$, to the posterior law of the impulse response. When the underlying system is deterministic and LTI, and inference is performed in a model class that enforces or symmetrically favors time-invariant impulse responses, this approximate posterior may be interpreted as an LTIE law induced by inference: the posterior mean is time-invariant, while the posterior spread captures uncertainty about the unknown filter rather than literal physical channel fluctuations. From this approximate posterior, we can then derive the uncertainty for any related quantity, such as the denoised latent signal or the cross-correlation function (CCF).

\subsubsection{Problem Formulation}

We consider a standard discrete-time LTI channel where a known, deterministic signal $f[n]$ is transformed by an unknown impulse response $h[k]$ to produce a latent noiseless output
\begin{equation}
g[n] = (f * h)[n],
\end{equation}
and the observed signal is
\begin{equation}
y[n] = g[n] + \nu[n],
\end{equation}
where $*$ denotes convolution and $\nu[n]$ is assumed to be zero-mean white Gaussian noise. This is the standard Bayesian observation model for LTI system identification, in which additive observation noise is modeled explicitly rather than being absorbed into a stochastic impulse-response model. In this subsection we use the conventional LTI indexing $h[k]$ for the FIR coefficients; this differs from the no-direct-feedthrough convention $k=1,\ldots,p$ used in Section~\ref{s:bayesian_framework} only by an index shift and has no substantive effect on the analysis.

A common analysis technique is to compute the cross-correlation between the input and output. We use the convention
\[
(a \otimes b)[n] \coloneqq \sum_{m=-\infty}^{\infty} a^*[m]\, b[m+n],
\]
so that $(f \otimes f)[n]$ is the autocorrelation of the input signal. Assuming $\nu[n]$ is zero-mean and uncorrelated with the deterministic input $f[n]$, we have $\mathbb{E}[(f \otimes \nu)[n]] = 0$, and therefore
\begin{equation}
\mathbb{E}[ (f \otimes y)[n] ] = (f \otimes f) * h[n].
\end{equation}
For finite-length signals, this identity holds up to the usual boundary conventions (e.g., zero-padding or valid-overlap indexing). Thus, in expectation, the input-output cross-correlation is the convolution of the input autocorrelation with the impulse response. Classically, one could attempt to estimate $h[n]$ by deconvolving the input autocorrelation from the observed CCF. However, direct deconvolution is often unstable and prone to noise amplification \citep{BerteroBoccacci2005}.

Our Bayesian approach avoids deconvolution entirely. Instead, we treat the impulse response $h$ as a vector of random variables and use the observed data pair $(f, y)$ to perform inference and find its posterior distribution.

\subsubsection{Experimental Setup and Method}

To create a realistic test case, the input signal $f[n]$ was taken from the publicly available "Earthquakes" dataset \citep{EarthquakesUCR2018}. In our visual inspection, the selected series exhibits non-stationary, pulse-like structure. A ground truth Finite Impulse Response (FIR) filter $h[k]$ was synthetically generated. The latent clean output was computed as $g[n] = (f * h)[n]$, and the final observed signal $y[n]$ was created by adding zero-mean Gaussian noise with standard deviation $0.5$.

The core of our method is to model the convolution operation using a Bayesian neural network. Specifically, the impulse response $h$ is parameterized through the kernel of a 1D Bayesian convolutional layer. We use the Flipout estimator \citep{Wen2018}, as implemented in the \texttt{tfp.layers.Convolution1DFlipout} layer in TensorFlow Probability \citep{tensorflow2015-whitepaper, Dillon2017TFD, TFPConvolution1DFlipoutDocs}, which offers a low-variance method for estimating gradients. Since this layer implements cross-correlation under the library convention, the impulse response reported in our plots is obtained by reversing the learned kernel coefficients before comparison with the ground truth. The layer learns a posterior distribution for its kernel weights.

\begin{itemize}
    \item \textbf{Model and Prior:} The network consists of a single Bayesian convolutional layer. Following common practice for regularization, we place a fixed, zero-mean Gaussian prior on the weights of the kernel. The standard deviation of this prior is set to $1/\sqrt{\text{kernel\_size}}$, which helps prevent overfitting by discouraging excessively large filter coefficients. The posterior distribution is approximated as a diagonal Gaussian (mean-field approximation), which is learned during training. As a result, the approximation cannot represent posterior correlations between nearby taps, so uncertainty propagated to derived quantities such as transfer functions and CCFs does not capture the full posterior dependence structure.

    \item \textbf{Loss Function and Inference:} We use variational inference to approximate the posterior. In practice, we minimize an ELBO-style variational objective \citep{TFPConvolution1DFlipoutDocs}. The ability to compute the gradient of the expectation term with respect to the variational parameters is enabled by the reparameterization trick \citep{Kingma2013, Rezende2014}. The corresponding ELBO consists of two terms:
    \begin{enumerate}
        \item The \textit{expected log-likelihood}, which measures how well the model's predictions fit the data. For an assumed Gaussian observation model, this term yields a squared-error data-fit term up to additive constants and noise-variance scaling.
        \item The \textit{Kullback-Leibler (KL) divergence}, which measures the "cost" of deviating from the prior. This term regularizes the model by penalizing overly complex posteriors.
    \end{enumerate}
    
The final loss used in practice is
\[
\text{Loss} = \lambda \, ||y - \hat{g}||_2^2 + \beta \cdot \text{KL}(q(h) || p(h)),
\]
where $q(h)$ is the approximate posterior, $p(h)$ is the prior, $\hat{g}$ is the noiseless model output induced by a sampled impulse response, and $\beta$ scales the KL divergence term according to the number of statistically distinct training examples represented in the objective. Under a Gaussian observation model with observation-noise standard deviation $\sigma_{\mathrm{obs}}$, the implementation uses $\lambda = 1/(2\sigma_{\mathrm{obs}}^2)$. In the single-observation experiment reported here, $\sigma_{\mathrm{obs}}=0.5$ was fixed rather than learned, so $\lambda$ remained constant throughout training. The current LTIE implementation also supports updating this variance scale from residuals through a non-trainable moving-average estimate outside backpropagation, but that option was not used in this experiment. In the single-observation experiment, the number of statistically distinct training examples is $1$, consistent with the TensorFlow Probability recommendation that the KL contribution be scaled so it is applied once per epoch \citep{TFPConvolution1DFlipoutDocs}. Under the fixed-$\sigma_{\mathrm{obs}}$ setting used here, this objective is proportional to the negative ELBO up to additive constants.

    \item \textbf{Training on a Single Observation:} To train the model on a single pair $(f,y)$, we employ a Monte Carlo technique. The single input is repeated to form a large computational batch (e.g., \texttt{batch\_size=1024}). During each training step, the network samples a different kernel $h$ from the current posterior approximation for each repeated instance in the batch. This provides a stable Monte Carlo estimate of the expectation in the ELBO's log-likelihood term, enabling effective gradient-based optimization even with a single data point. These repeated batch elements are used only for Monte Carlo estimation and do not constitute statistically distinct observations; accordingly, they do not affect the KL normalization. The model was trained using the Adam optimizer with a cosine decay learning rate schedule \citep{Loshchilov2017}. In experiments with multiple independent observation pairs, the KL term was normalized by the number of distinct pairs rather than by the computational batch size.
\end{itemize}

\subsubsection{Results and Discussion}
\begin{figure}[!ht]
\centering
\includegraphics[width=\linewidth]{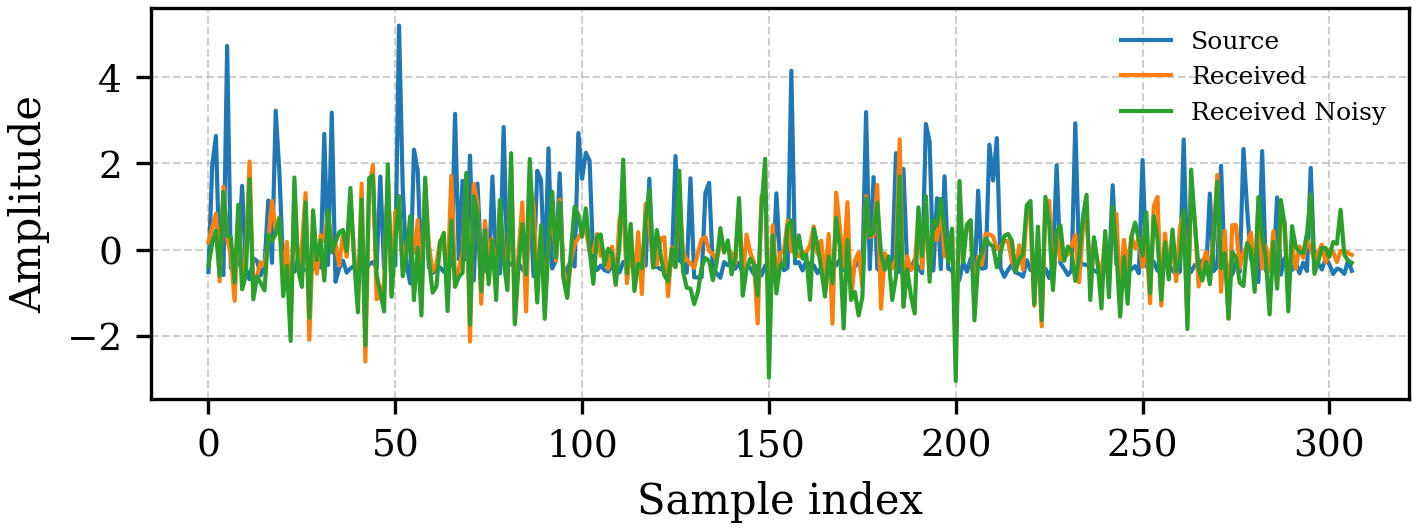}
\caption{The synthetic data used for experiment one- impulse response regression from a single observation.}
\label{fig:experiment1_a_all_data}
\end{figure}

\begin{figure}[!ht]
    \centering
    \begin{subfigure}[b]{.47\linewidth}
        \includegraphics[width=\linewidth]{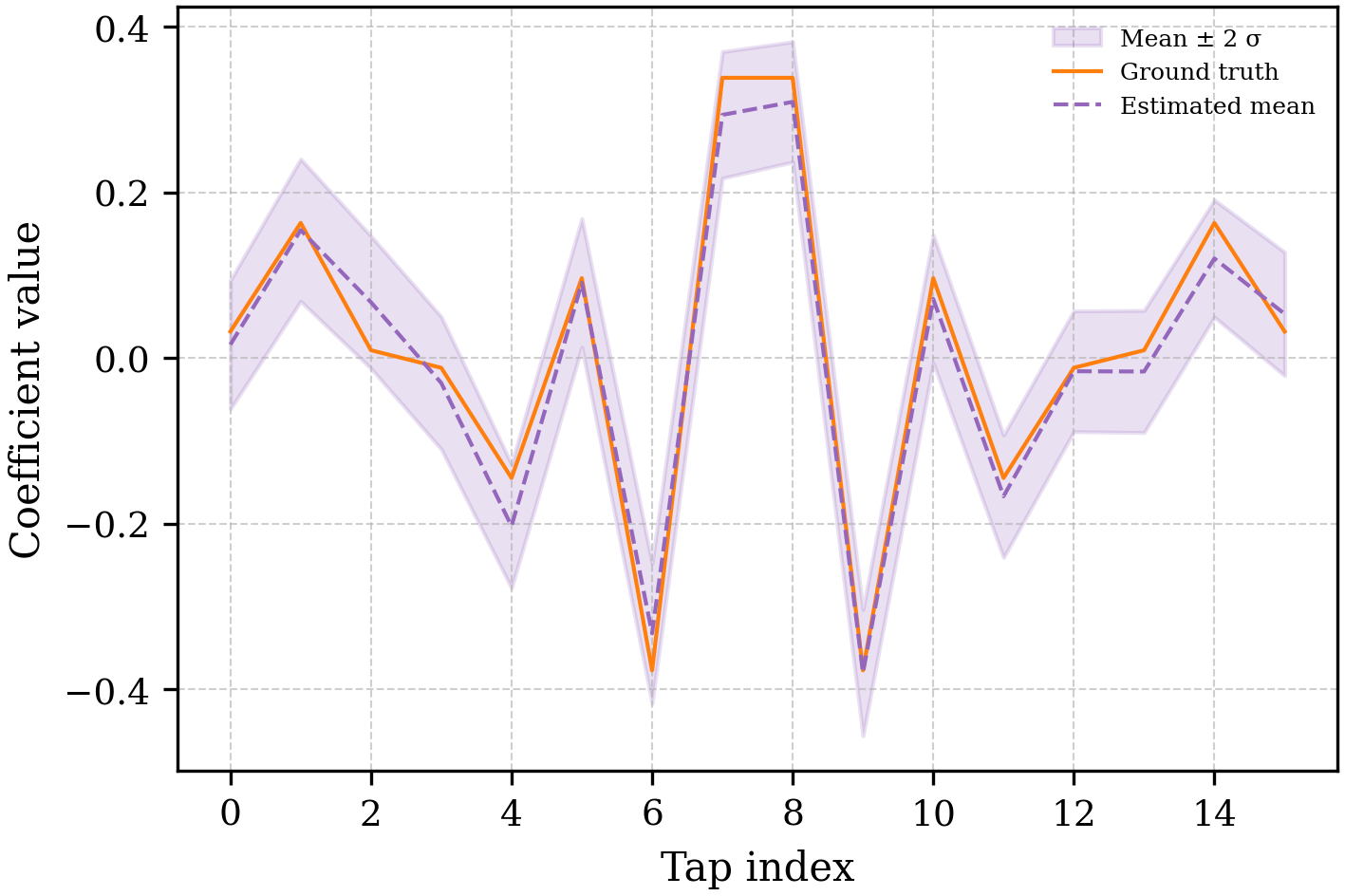}
        \subcaption{The posterior mean (blue dashed line) of the estimated impulse response tracks the ground truth (orange). The shaded region shows the posterior mean $\pm 2$ posterior standard deviations.}
        \label{fig:experiment1_a_samples_posterior_ir_estimate}
    \end{subfigure}
    \hfill
    \begin{subfigure}[b]{.47\linewidth}
        \includegraphics[width=\linewidth]{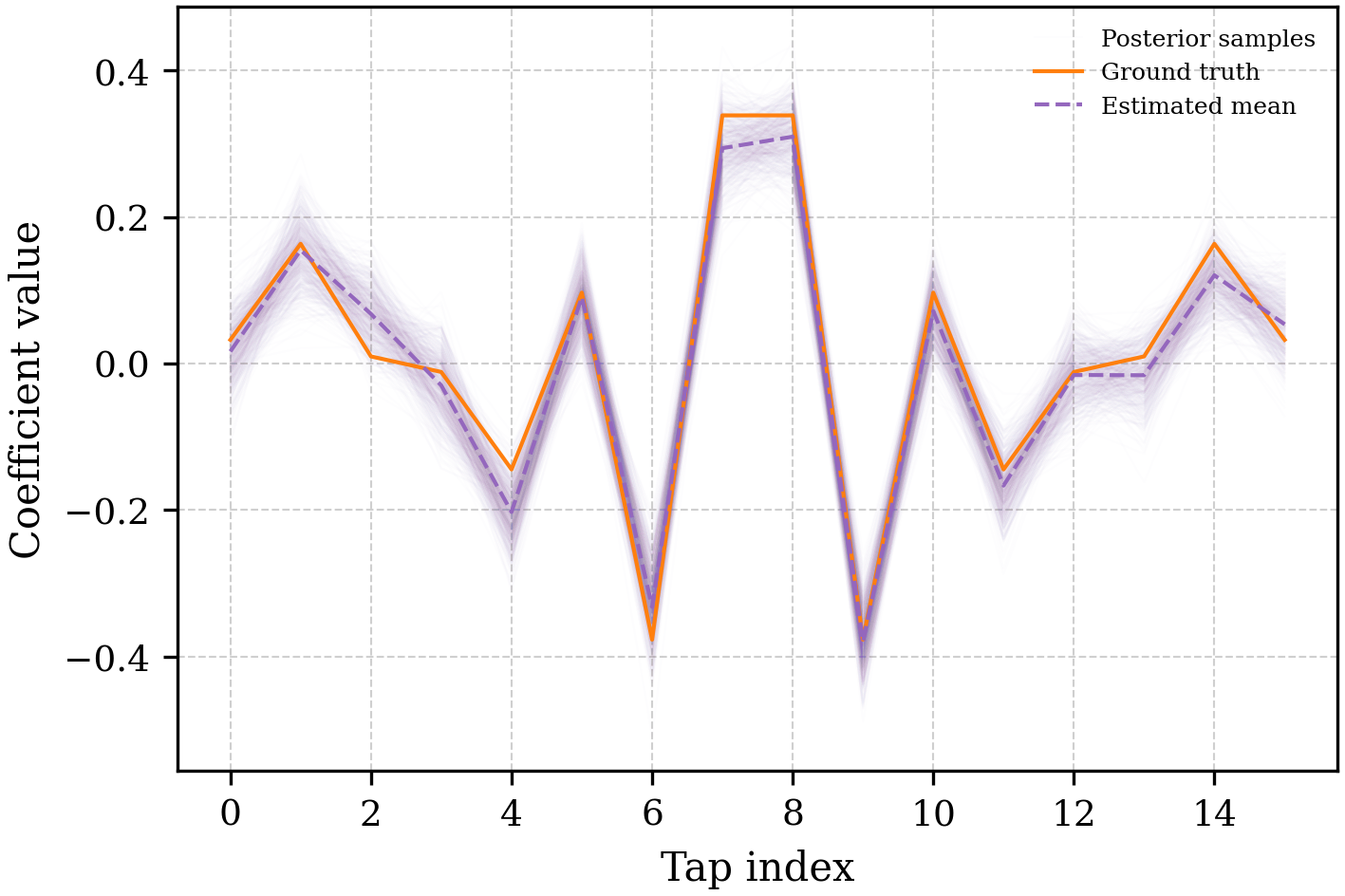}
        \subcaption{Individual samples drawn from the posterior distribution (blue lines), illustrating the model's uncertainty around the ground truth impulse response (orange).}
        \label{fig:experiment1_a_samples_posterior_ir_samples}
    \end{subfigure}
    \caption{Bayesian estimation of the impulse response, $h[k]$, from a single noisy observation. The model learns a full posterior distribution that successfully recovers the ground truth filter and quantifies the associated estimation uncertainty.}
\end{figure}

The results demonstrate a successful recovery of the system's properties from a single noisy observation pair (visualized in Figure~\ref{fig:experiment1_a_all_data}). After training, the model yields a variational posterior distribution for the impulse response $h$. As shown in the appendix, the posterior mean correctly tracks the ground truth filter, and the posterior variance provides a useful uncertainty summary (see Figures~\ref{fig:experiment1_a_samples_posterior_ir_estimate} and \ref{fig:experiment1_a_samples_posterior_ir_samples}). A repeated-trial pointwise coverage check for the approximate variational posterior showed conservative intervals in this synthetic LTIE setting: nominal $90\%$ and $95\%$ credible intervals achieved empirical coverages of $97.5\%$ and $99.4\%$ for the single-observation case, and remained conservative while contracting as additional independent observation pairs were incorporated (\ref{app:experimental_settings}).

\begin{figure}[!ht]
\centering
\includegraphics[width=\linewidth]{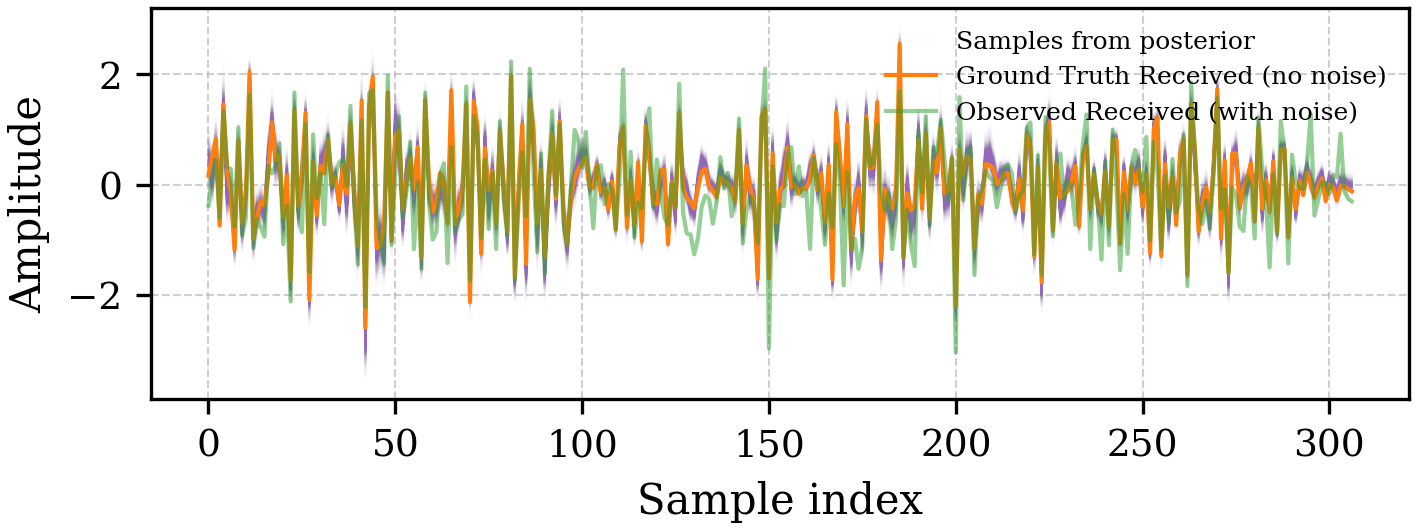}
\caption{Denoising performance of the Bayesian model in the single-observation experiment. The figure shows the posterior predictive distribution for the received signal, $\hat{g}[n]$ (a collection of green samples). By learning the impulse response, the model generates predictions that accurately reconstruct the clean signal while filtering the noise from the original observation.}
\label{fig:denoising_performance} 
\end{figure}

\begin{figure}[!ht]
    \centering
        \includegraphics[width=\linewidth]{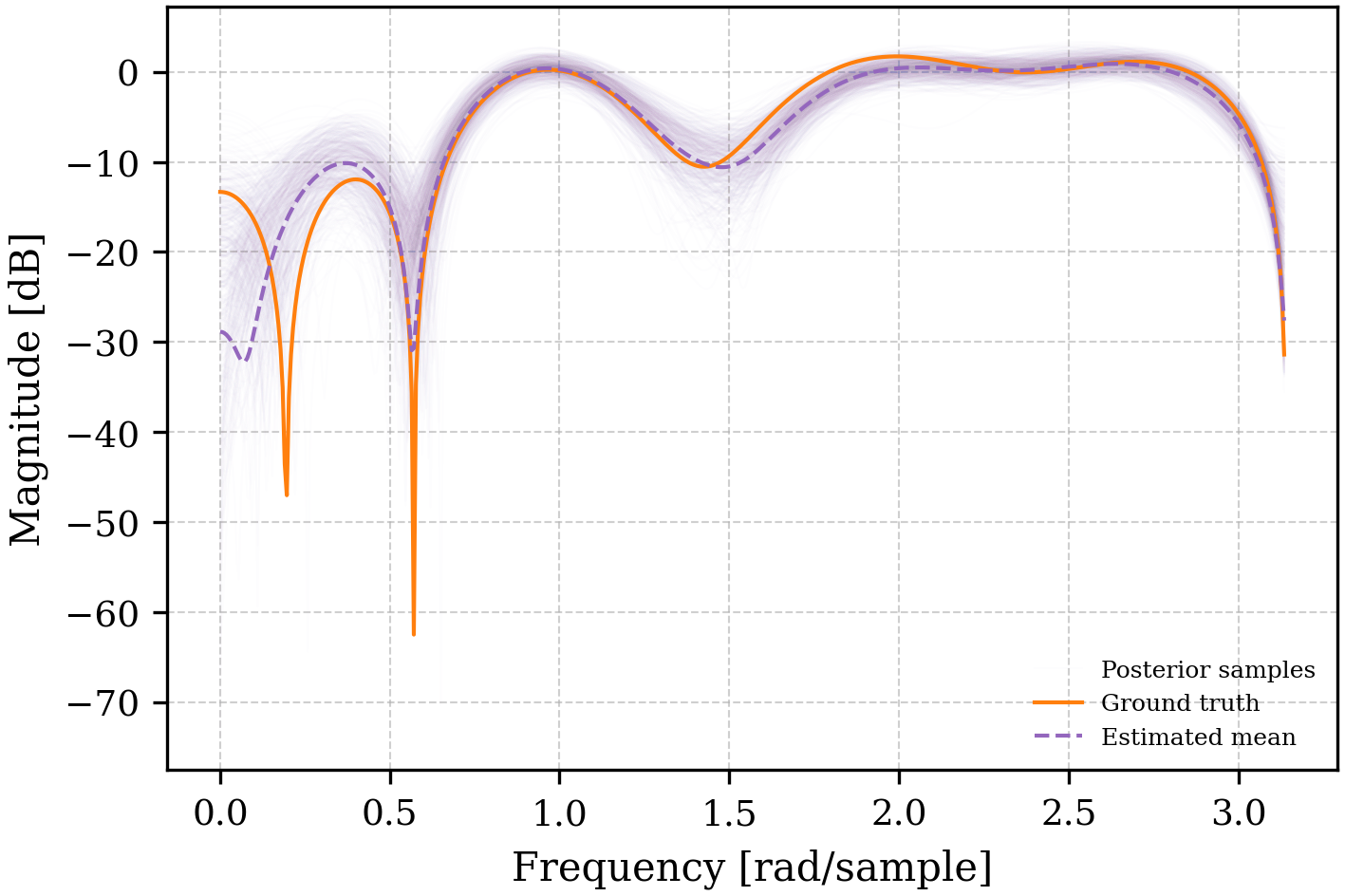}
        
    \caption{Propagation of uncertainty into the frequency domain. The variational posterior distribution over the impulse response induces a corresponding approximate posterior over the system's frequency and phase response, allowing for comprehensive uncertainty analysis.}
    \label{fig:experiment1_a_samples_posterior_freq}
\end{figure}

\begin{figure}[!ht]
    \centering
    \begin{subfigure}[b]{0.9\textwidth}
        \includegraphics[width=\linewidth]{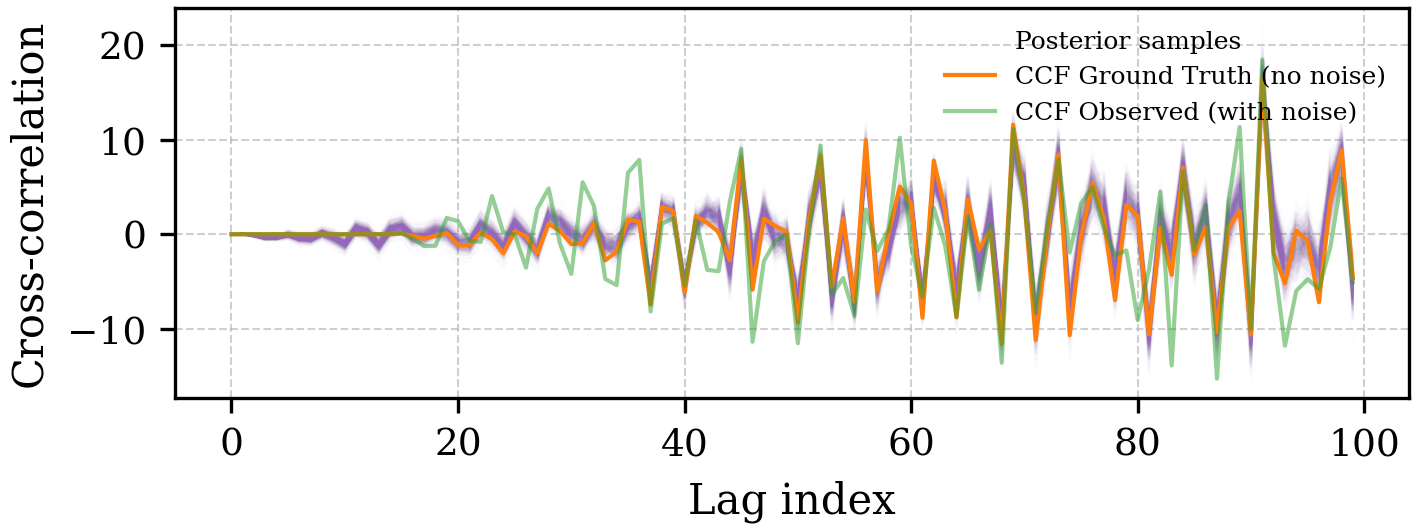}
        \subcaption{Zoom on lags near 0.}
        \label{fig:app:xcorr_zoom0}
    \end{subfigure}

    \vspace{1em}

    \begin{subfigure}[b]{0.9\textwidth}
        \includegraphics[width=\linewidth]{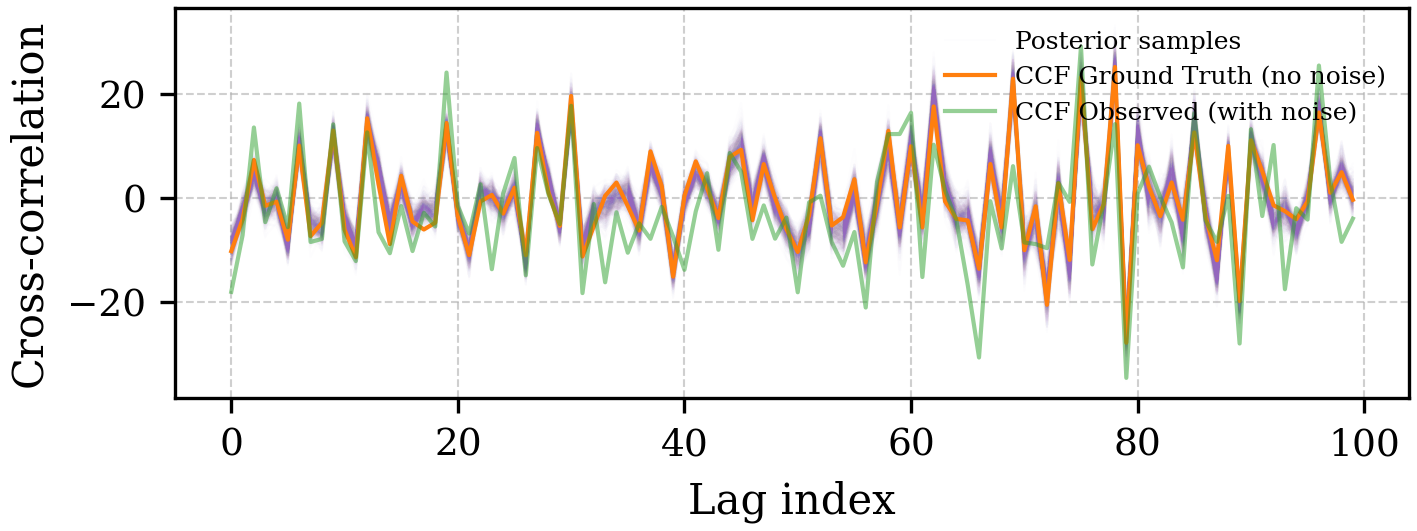}
        \subcaption{Zoom on lags near 100.}
        \label{fig:app:xcorr_zoom100}
    \end{subfigure}

    \vspace{1em}

    \begin{subfigure}[b]{0.9\textwidth}
        \includegraphics[width=\linewidth]{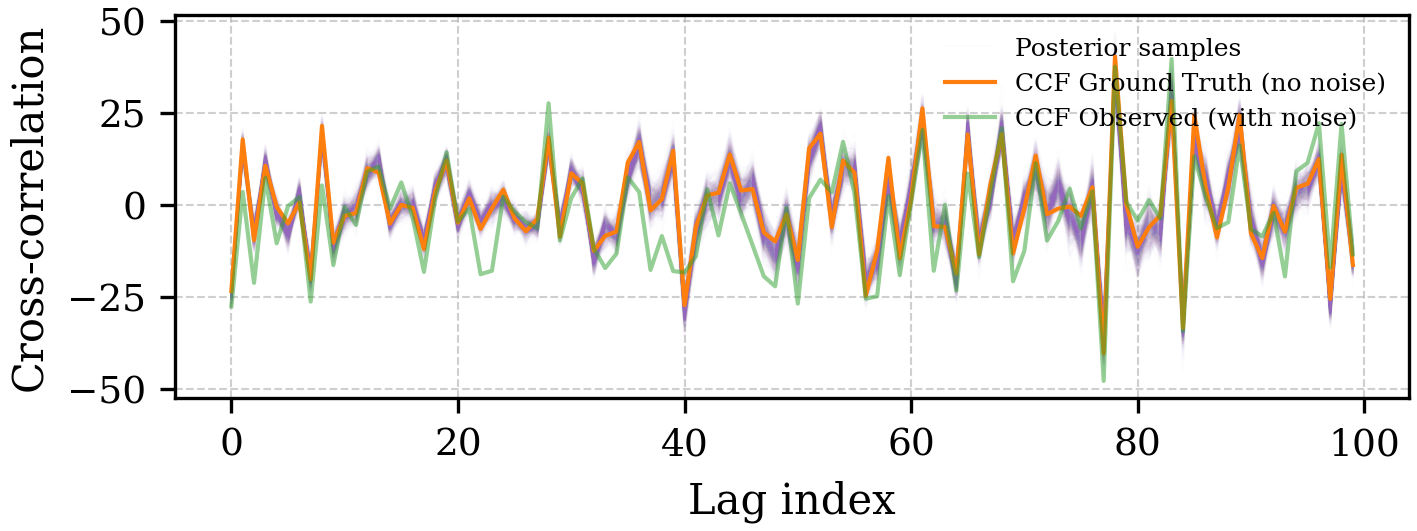}
        \subcaption{Zoom on lags near 200.}
        \label{fig:app:xcorr_zoom200}
    \end{subfigure}

    \vspace{1em}

    \begin{subfigure}[b]{0.9\textwidth}
        \includegraphics[width=\linewidth]{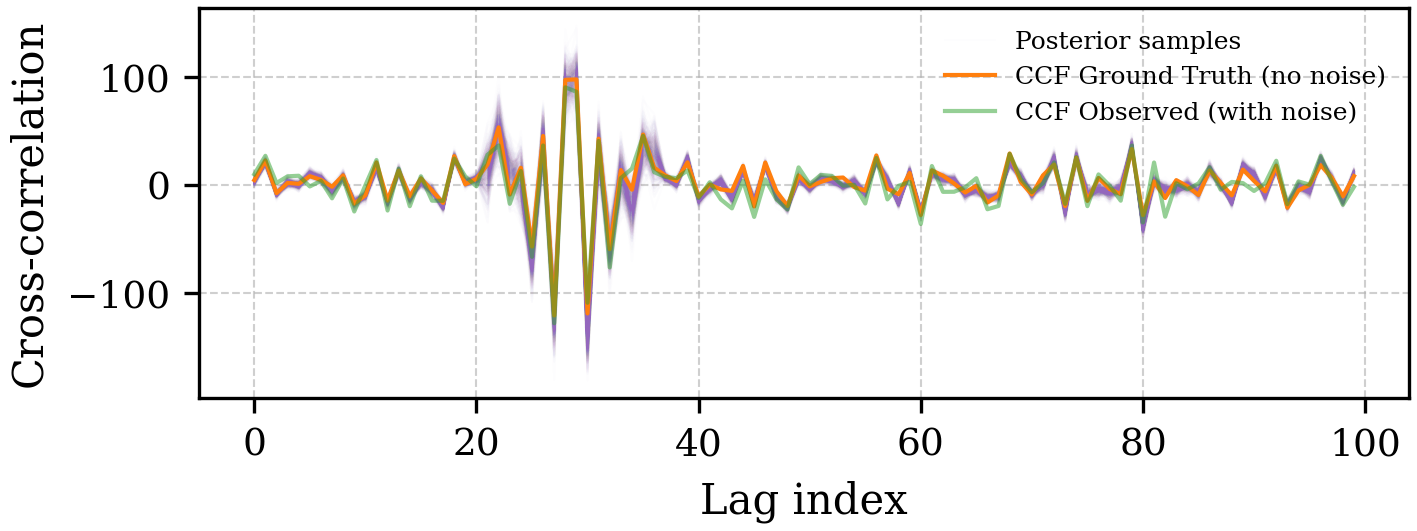}
        \subcaption{Zoom on lags near 300.}
        \label{fig:app:xcorr_zoom300}
    \end{subfigure}
    
    \caption{Samples from the posterior distribution of the cross-correlation function (CCF) for the single-observation experiment. These plots show the estimated CCF distribution (green) robustly matching the ground truth (orange) while providing a credible interval, in contrast to the noisy observed CCF (red). Each panel shows a zoomed-in view of a different lag region.}
    \label{fig:app:xcorr_all} 
\end{figure}

A key advantage of the Bayesian approach is the ability to propagate this learned uncertainty to any derived quantity. For instance, by passing the input signal through posterior samples of the filter, we can generate a full posterior distribution for the denoised signal, $\hat{g}[n]$. Figure~\ref{fig:denoising_performance} demonstrates that these predictions successfully reconstruct the clean ground truth signal, effectively filtering the observation noise. This uncertainty propagation extends to the frequency domain, where the posterior over $h[k]$ induces a posterior over the system's transfer function (Figure~\ref{fig:experiment1_a_samples_posterior_freq}). Finally, the framework can robustly estimate the cross-correlation function (CCF) with credible intervals, a significant improvement over a single, noisy CCF calculated from the observation, as shown in Figure~\ref{fig:app:xcorr_all}.

This experiment validates that the Bayesian inference machinery at the heart of our framework can be practically applied to solve classic signal processing challenges, turning an ill-posed deconvolution problem into a well-behaved probabilistic inference task. 

While this demonstration focuses on the challenging single-observation scenario, it is important to note the model's behavior with additional data. In further tests where the model was conditioned on multiple independent observation pairs, the posterior distribution of the impulse response systematically tightened, reflecting an increase in confidence and a reduction in epistemic uncertainty as more evidence was incorporated.

Accordingly, the success of this experiment should be interpreted as evidence of robustness under model misspecification and of the usefulness of the posterior representation, rather than as evidence that the underlying data-generating mechanism is itself a stochastic impulse-response channel. From this perspective, the experiment also illustrates a conceptual contrast with the standard additive-noise model: the posterior law over the impulse response provides a useful uncertainty-aware representation even when the true data-generating mechanism includes a separate observation-noise term.

\subsection{Application to Simulated Ambient Noise Tomography}

This experiment demonstrates the framework's application to a problem inspired by geophysical Ambient Noise Tomography (ANT), where the goal is to recover the dispersive properties of a medium from ambient-noise cross-correlations \citep{Bensen2007,Ekstrom2009}. The experiment directly compares our Bayesian method against a classical frequentist approach (CCF stacking) on a simulated LTIE (Linear Time-Invariant in Expectation) system, where a stable impulse response is probed by a superposition of many random, independent pulses, mimicking a diffuse noise field.

\subsubsection{Methodology}
A synthetic ambient noise field was generated using a ground-truth dispersive velocity curve. The simulation is constructed so that, after ensemble averaging or sufficiently long stacking, the inter-station cross-correlation is proportional to the target Green's function over the band of interest, making comparison against an impulse-response estimate meaningful in this controlled setting \citep{Snieder2004, SanchezSesma2006}. The resulting signals were processed using two different methods:

\begin{enumerate}
    \item \textbf{Classical CCF Stacking:} Signal pairs were preprocessed with spectral whitening, segmented into windows, and cross-correlated; the resulting correlations were then averaged (stacked) to produce a single empirical CCF, following the standard ANT workflow \citep{Bensen2007}. This stacked CCF serves as an estimate of the impulse response.

    \item \textbf{Bayesian Regression:} Our model operated directly on the raw (non-whitened) signal pairs. It was used to deconvolve the source characteristics and infer an approximate posterior distribution for the true impulse response. The mean of this approximate posterior (Mean Impulse Response, or MIR) serves as our final estimate. The variational objective used the same KL normalization convention as in the previous experiments, namely scaling with the number of statistically distinct training pairs rather than the computational batch size. In this ANT experiment, the Gaussian data-fit term used a residual-estimated observation-noise scale, so the weighting $\lambda = 1/(2\hat{\sigma}_{\mathrm{obs}}^2)$ was updated during training via a non-trainable moving-average estimate outside backpropagation.
\end{enumerate}

For both methods, the final step was to derive the phase velocity curve from the estimated impulse response. This was achieved by fitting the frequency-domain representation of the estimate against theoretical beam patterns derived from the zeroth-order Bessel function of the first kind ($J_0$), following the standard spectral formulation used in ANT \citep{Aki1957, Ekstrom2009}.

Accordingly, this ANT comparison should be interpreted as a comparison between practical end-to-end pipelines rather than as a strictly preprocessing-matched estimator comparison. The distinction matters because the classical CCF baseline benefits from whitening for phase-based recovery, whereas our Bayesian method is designed to operate on raw data and to retain spectral-amplitude information relevant to impulse-response inference.

\subsubsection{Results and Discussion}
\begin{figure}[!htp]
    \centering
    \begin{subfigure}{0.75\textwidth}
        \centering
        \includegraphics[width=\linewidth]{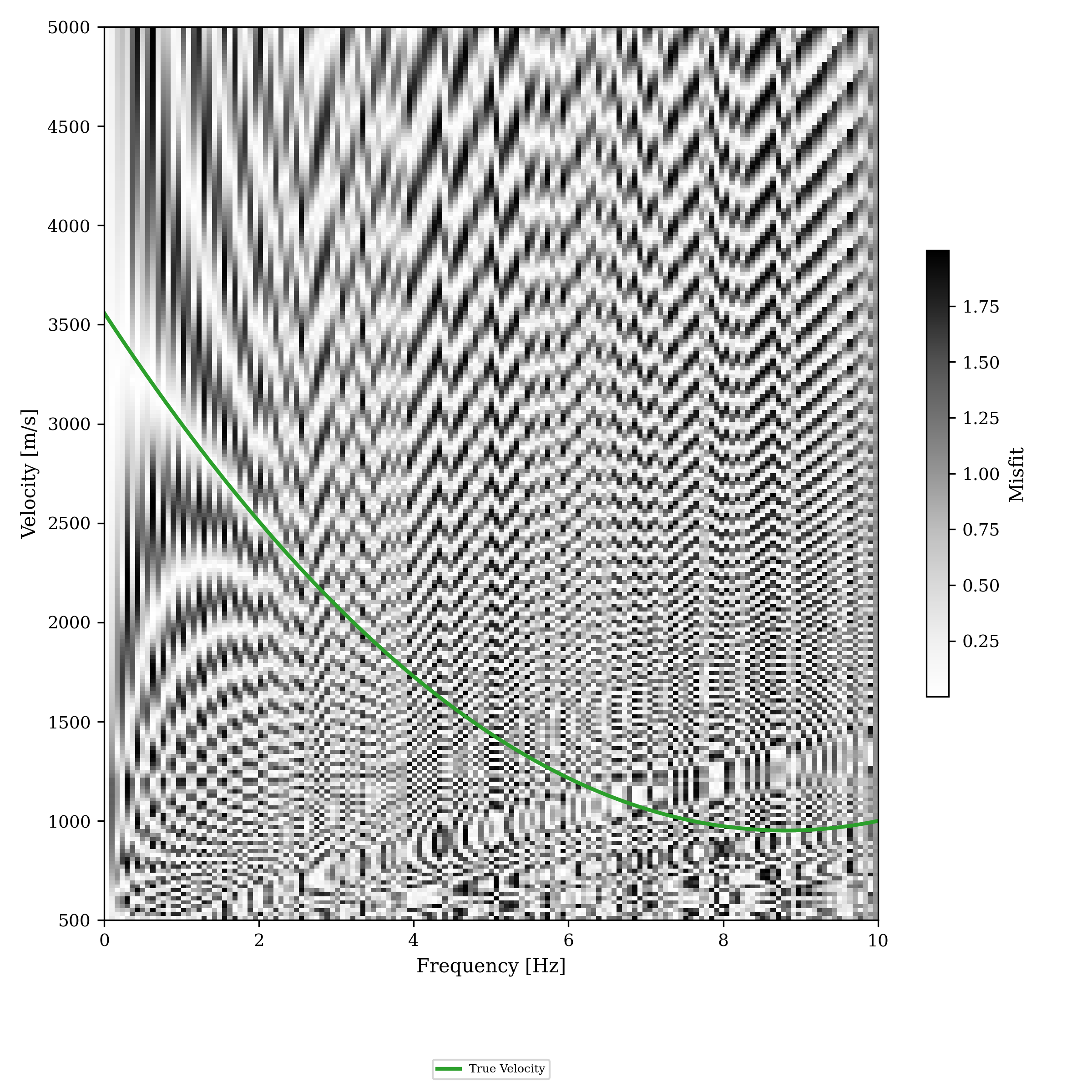}
        \caption{Velocity misfit map for the classical CCF estimate. The green line is the ground-truth velocity.}
        \label{fig:ant_misfit_map_ccf}
    \end{subfigure}

    \medskip 

    \begin{subfigure}{0.75\textwidth}
        \centering
        \includegraphics[width=\linewidth]{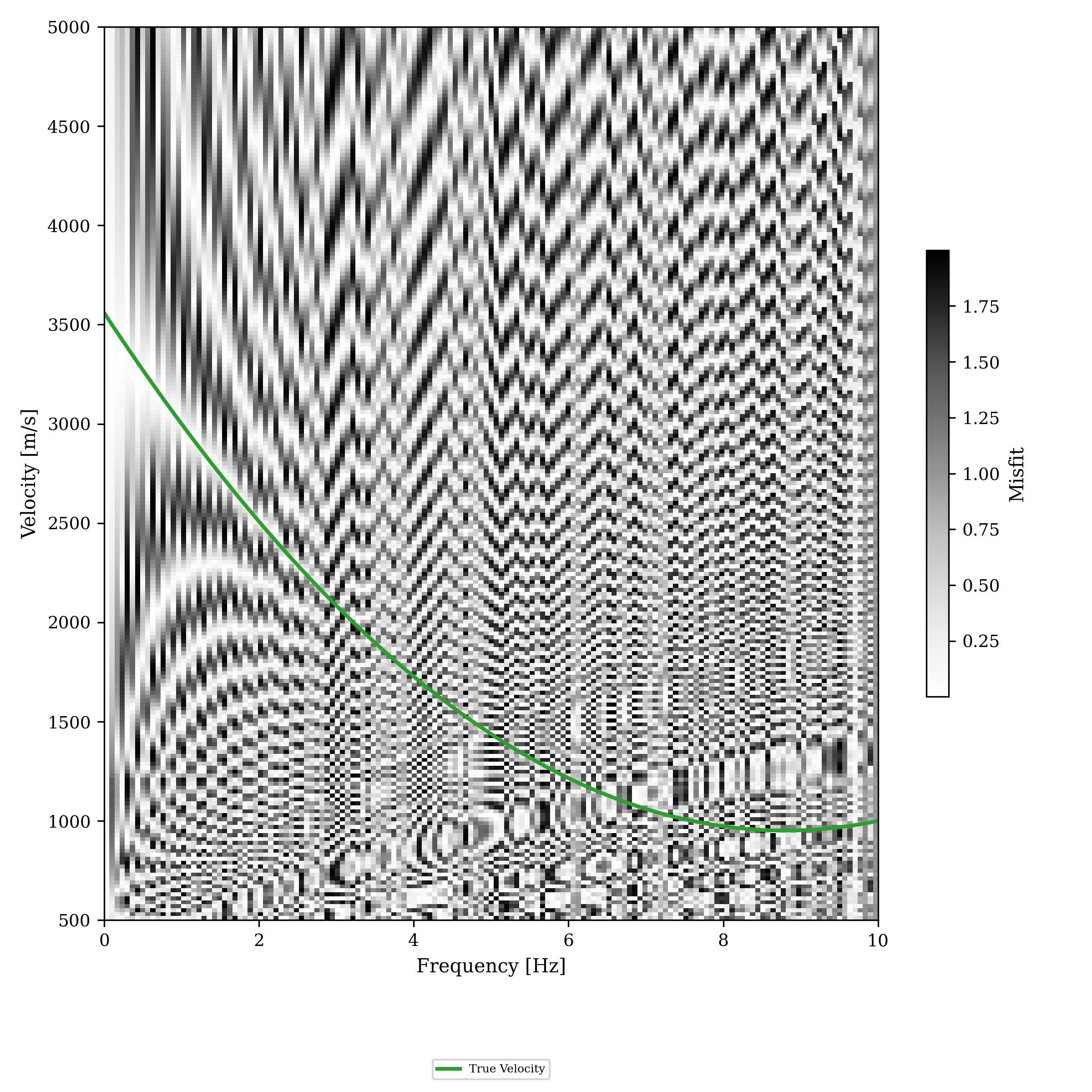}
        \caption{Velocity misfit map for the Bayesian MIR estimate. The green line is the ground-truth velocity.}
        \label{fig:ant_misfit_map_mir}
    \end{subfigure}

    \caption{Comparison of velocity misfit maps for (a) the classical CCF and (b) the Bayesian MIR methods. The misfit (error) is shown in greyscale, where darker shades indicate higher error. The Bayesian MIR shows a lower error margin around the ground-truth velocity curve (green), particularly at lower frequencies, and its low-misfit valley extends more smoothly into the higher frequency range (up to 3Hz). This leads to a more accurate and robust final velocity estimation.}
    \label{fig:ant_misfit_maps_comparison}
\end{figure}

Both methods effectively recovered the underlying velocity structure, but in these simulations the Bayesian approach showed a lower error floor. The reason for this improved performance is evident in the velocity misfit maps (Figure~\ref{fig:ant_misfit_maps_comparison}), where the Bayesian method produces a clearer result.

\begin{figure}[!h]
    \centering
    \includegraphics[width=0.8\textwidth]{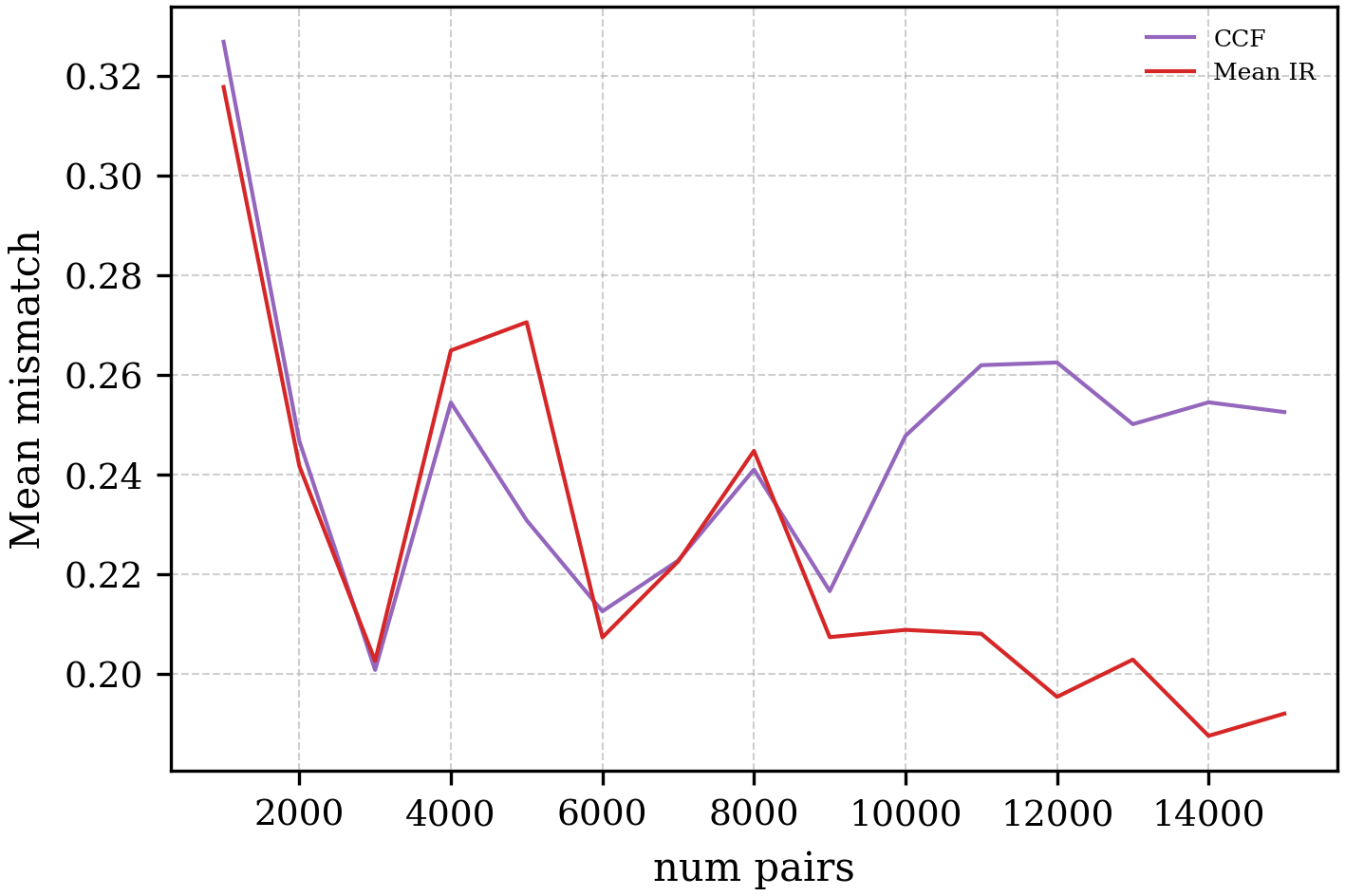}
    \caption{Performance comparison showing target velocity mismatch error as a function of the number of signal pairs used. The Bayesian MIR method (red) converges to a lower overall error floor compared to the classical CCF method (purple).}
    \label{fig:ant_error_vs_num_pairs}
\end{figure}

In the standard experiment (without quantization), our Bayesian model (MIR) shows better performance than the classical CCF method in this simulated setting, approaching a lower error floor in Figure~\ref{fig:ant_error_vs_num_pairs}, which plots the estimation error as a function of the number of pairs. Over the sweep from 1000 to 15{,}000 pairs, the MIR achieved a lower mean target-curve error (0.225 versus 0.245 for CCF) and a lower minimum error (approximately 0.188 versus 0.201 for CCF). In a representative 12{,}000-pair run, the MIR mean error using the recovered velocity curve was approximately 0.207, compared with 0.262 for the classical method.

A similar trend was observed in a more challenging scenario involving aggressive 1-bit quantization of the signals. While both methods remained effective, our Bayesian model again showed lower error in these simulations. Using the same settings except for doubling the Bayesian batch size, the MIR achieved a lower mean target-curve error across the 1000 to 15{,}000 pair sweep (0.241 versus 0.269 for CCF) and a lower minimum error (approximately 0.178 versus 0.221 for CCF). With 15{,}000 pairs, the MIR mean error at the true velocity was approximately 0.191, while the classical method's error was 0.242.

The results suggest that by learning a complete probabilistic model of the impulse response rather than simply averaging correlations, our method can extract system properties more accurately and from fewer observations in this simulated setting. This is particularly advantageous in data-limited scenarios and highlights the practical benefits of the Bayesian framework. A significant qualitative advantage of our approach is that it does not require the spectral whitening step common in classical ANT pipelines \citep{Bensen2007}. While whitening often improves the quality of Green's function estimates in the classical phase-focused pipeline, it also alters the original spectral amplitudes of the transfer function. By working with the raw data, our method retains the system's frequency response, allowing for an estimation of the transfer function between the virtual source and receiver. This task is more challenging in classical pipelines where the effects of whitening are difficult to reverse.

\subsection{Regression of a Non-Stationary Impulse Response}

Building on the inference of time-invariant systems, this experiment addresses the more challenging problem of regressing a continuously changing Linear Time-Varying (LTV) impulse response. We show how a neural amortized variational inference model, with a deterministic CNN encoder and a Bayesian latent impulse-response layer endowed with a structured Gaussian Process (GP) prior, can effectively regularize this ill-posed problem, allowing for the robust estimation of a continuously varying impulse response from a single, noisy observation pair.

\subsubsection{Formulation and Experimental Setup}

\begin{figure}[!ht]
    \centering
    \includegraphics[width=\linewidth]{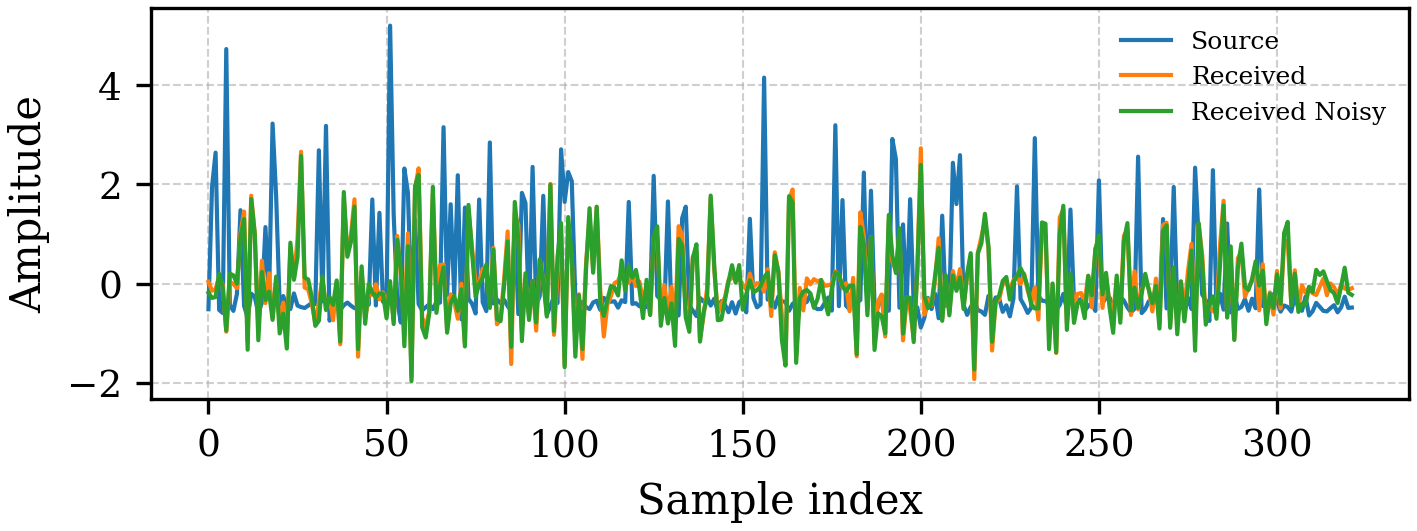}
    \caption{Synthetic data for the LTV system identification experiment. The figure shows: the input signal $f[n]$, the ground-truth time-varying impulse response $h[n, k]$, and the resulting noisy observed output signal $y[n]$.}
    \label{fig:ltv_experiment1_example_pair}
\end{figure}

\begin{figure}[!ht]
    \centering
    \begin{subfigure}[b]{.47\linewidth}
        \centering
        \includegraphics[height=3.8cm, keepaspectratio]{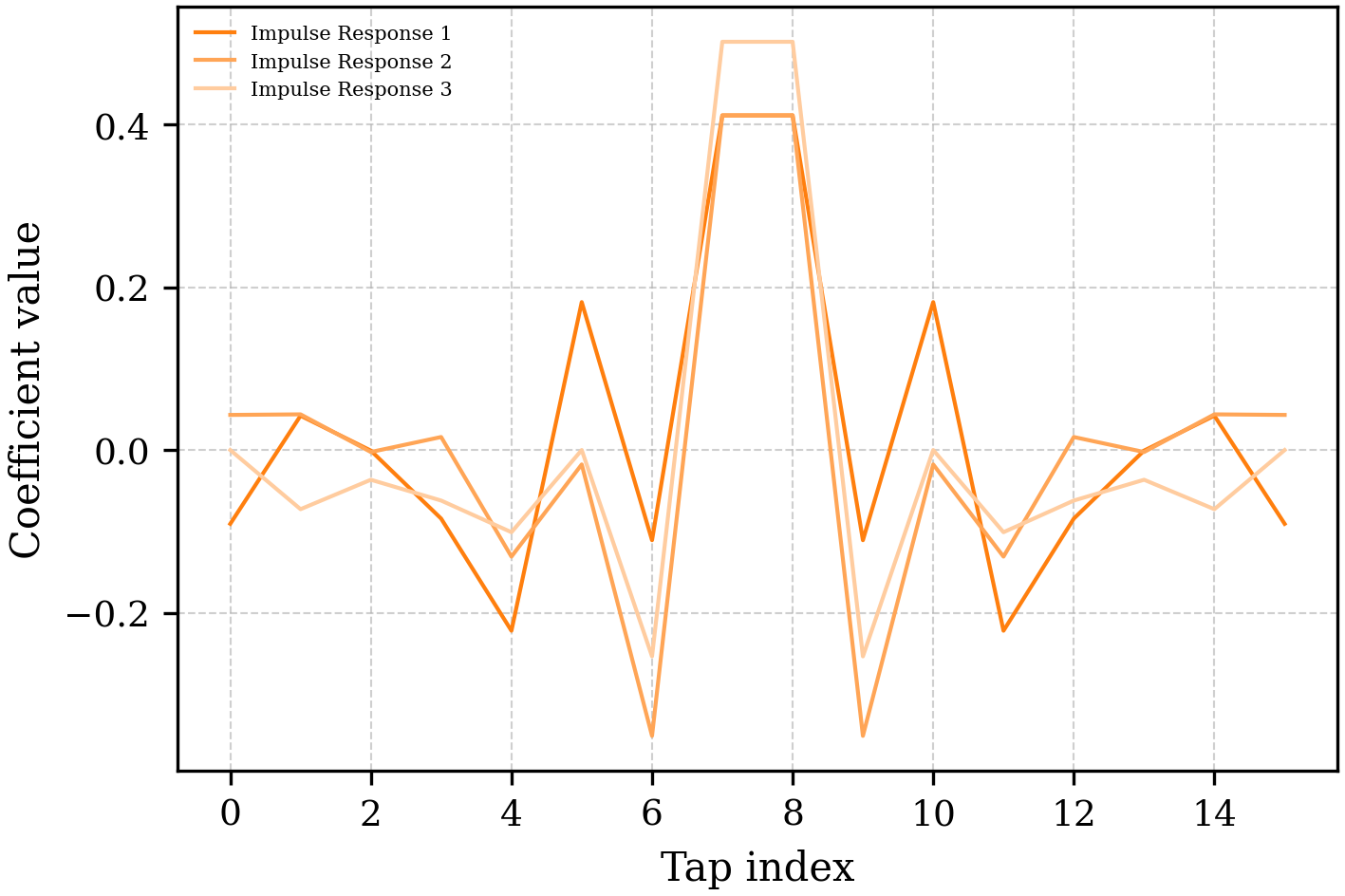}
        \subcaption{The three distinct time-invariant FIR filters that serve as the basis for the LTV system.}
        \label{fig:ltv_experiment1_three_fir_components}
    \end{subfigure}
    \hfill
    \begin{subfigure}[b]{.47\linewidth}
        \centering
        \includegraphics[height=4cm, keepaspectratio]{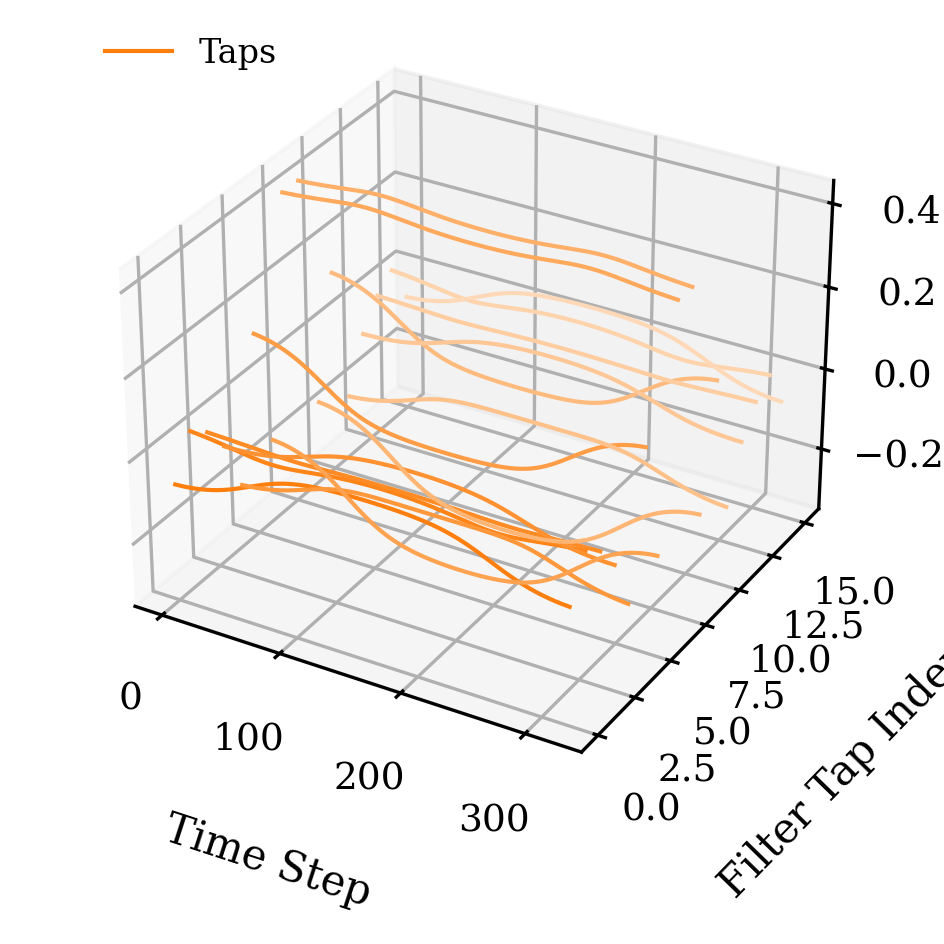}
        \subcaption{The final time-varying impulse response created by smoothly interpolating between the three base filters over time.}
        \label{fig:ltv_experiment1_ground_truth_ltv_3d}
    \end{subfigure}
    \caption{Generation of the ground-truth time-varying impulse response, $h[n, k]$. The LTV system is synthesized by smoothly interpolating between the three distinct FIR components shown in (a) to produce the final time-varying response shown in (b).}
    \label{fig:ltv_ground_truth_generation}
\end{figure}

To simulate a non-stationary channel, a time-varying impulse response $h[n,k]$ was created by smoothly interpolating between three distinct Finite Impulse Responses (FIRs), as shown in Figure~\ref{fig:ltv_ground_truth_generation}. The interpolation is governed by a set of time-varying weights $\alpha_i[n]$:
\begin{equation}
h[n] = \alpha_1[n] h^{(1)} + \alpha_2[n] h^{(2)} + \alpha_3[n] h^{(3)},
\end{equation}
where $\alpha_i[n] \ge 0$ and $\sum_{i=1}^{3} \alpha_i[n] = 1$ for all $n$, and $h[n]$ is the impulse response vector at time $n$. A single seismic signal from the "Earthquakes" dataset was used as the known input signal $f[n]$. The latent noiseless output $g[n]$ was generated by convolving $f[n]$ with the time-varying $h[n,k]$, and the observed output was then defined as $y[n] = g[n] + \nu[n]$ with white Gaussian noise $\nu[n]$. The complete set of synthetic signals used is shown in Figure~\ref{fig:ltv_experiment1_example_pair}. It is important to note that this generative model, with additive observation noise, differs from the latent stochastic impulse-response model developed earlier. This mismatch is intentional: the experiment is designed as a robustness test to assess whether the posterior representation remains useful even when the assumed source of uncertainty differs from the true data-generating mechanism.

\subsubsection{Modeling and Inference}

The core of the problem is to estimate the unknown $h[n,k]$ given the single observation pair $(f, y)$. This is ill-posed, as there are more unknown values in $h$ than observations at any given moment. To make this tractable, we regularize the problem by modeling the impulse response not point-by-point, but over a \textit{window of time} as a draw from a truncated Gaussian Process prior (specifically, over 32 time steps).

To perform inference, we use amortized variational inference, implemented with TensorFlow Probability. Rather than using a weight-space Bayesian Neural Network, we use a deterministic Convolutional Neural Network (CNN) as the inference encoder, and its outputs parameterize the approximate posterior distribution of a Bayesian latent impulse-response layer. The single long observation pair is partitioned into smaller, overlapping training examples. The CNN learns to map an input window directly to the parameters of the approximate posterior distribution for the corresponding impulse response window, $q(h_{\text{window}})$. This amortizes the cost of inference across all local windows, creating an efficient and scalable model. Because neighboring windows overlap, however, these training examples are not statistically independent observations of the full problem; accordingly, the objective below should be interpreted as a practical localized variational surrogate rather than as an exact global ELBO for an i.i.d. dataset.

To encourage smoothness and temporal consistency, we define a structured GP prior, $\mathcal{GP}_{\text{prior}}$, assuming each tap evolves smoothly over time according to a squared exponential (RBF) kernel. In particular, the prior covariance is block-diagonal across taps, so that $k((n,k),(m,\ell)) = k_t(n,m)\,\delta_{k\ell}$ with an RBF kernel $k_t$ over time and no prior coupling between distinct taps. The model is trained by minimizing a localized negative ELBO-style objective. Specifically, the loss for a given window is:
\begin{equation}
\mathcal{L} = \mathbb{E}_{q(h_{\text{window}})} \left[ || y_{\text{window}} - \hat{g}_{\text{window}} ||_2^2 \right] + \beta \cdot \text{KL}(q(h_{\text{window}}) || \mathcal{GP}_{\text{prior}}),
\end{equation}
where the expectation is approximated via Monte Carlo sampling. To reduce estimator variance, a different sample of the impulse response window is drawn from the posterior for each training example in the batch, providing a stable estimate of the expected loss. The first term is the expected L2 error, and the second is the KL divergence between the approximate posterior $q$ and the smooth GP prior. Under a Gaussian observation model, the first term is proportional to a negative expected log-likelihood up to additive constants and a fixed noise-variance scaling. Optimization is performed using the Adam optimizer with a cosine learning schedule.

\subsubsection{Results and Discussion}

\begin{figure}[!ht]
    \centering
    \begin{subfigure}[b]{.47\linewidth}
        \includegraphics[width=\linewidth]{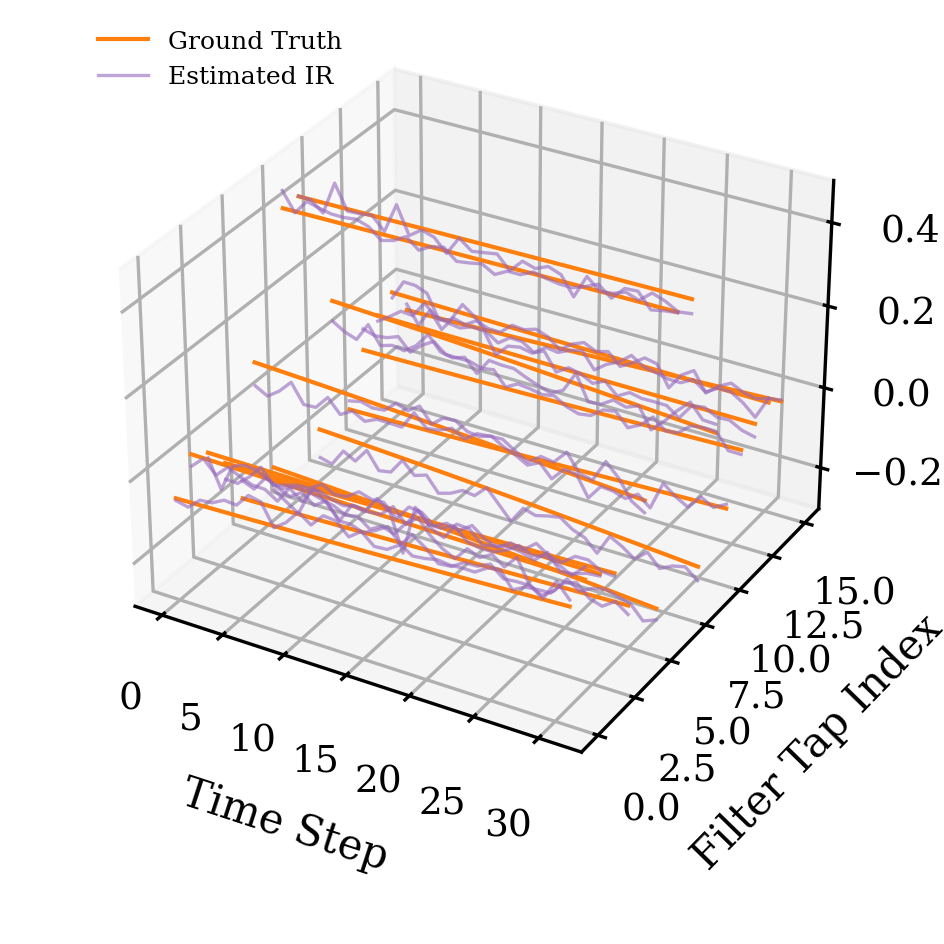}
        \subcaption{The estimated FIR (purple) and ground truth (orange) for a single time window. The smoothness of the estimate within this window is regularized by the Gaussian Process prior.}
        \label{fig:ltv_experiment1_single_window_fit}
    \end{subfigure}
    \hfill
    \begin{subfigure}[b]{.47\linewidth}
        \includegraphics[width=\linewidth]{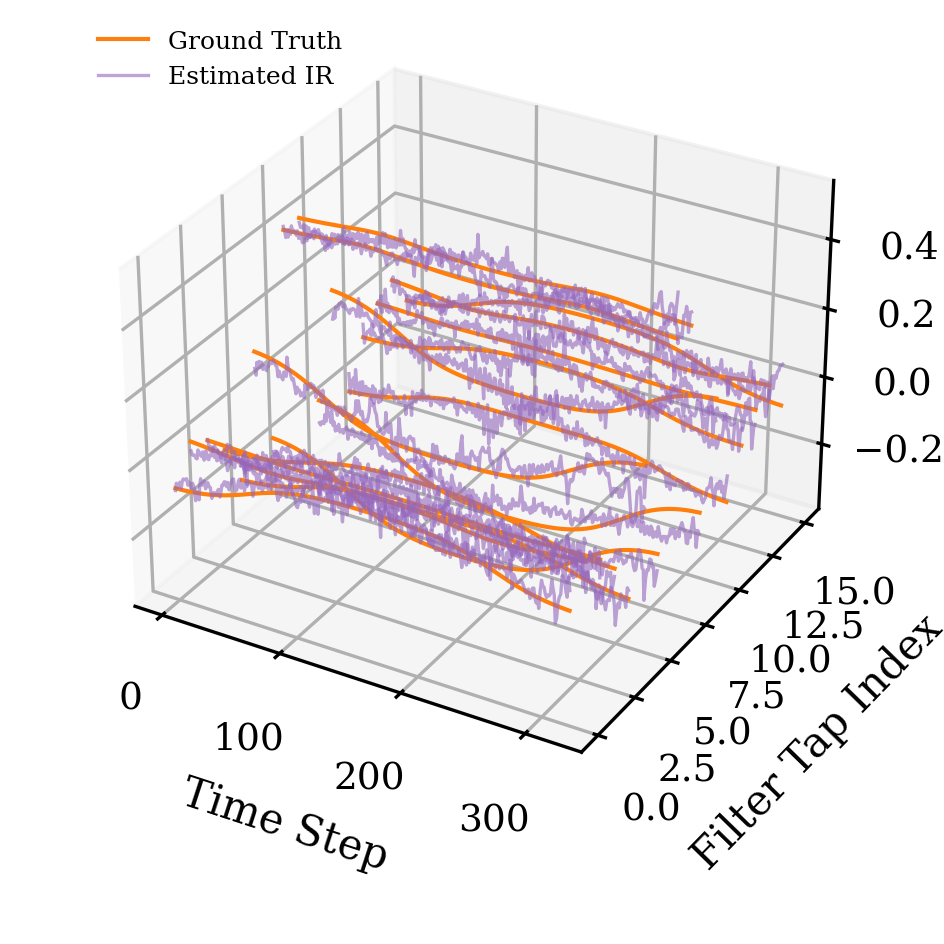}
        \subcaption{The complete LTV impulse response estimate (purple), created by averaging the predictions from all overlapping windows, shown against the ground truth (orange).}
        \label{fig:ltv_experiment1_stitched_fit}
    \end{subfigure}
    \caption{Estimation of the time-varying impulse response, $h[n, k]$. (a) The model's estimate for a single time window. (b) The complete LTV response, reconstructed by stitching together the estimates from all overlapping windows. The final result successfully tracks the ground-truth system dynamics.}
\end{figure}

Despite being trained on a single time series, the model was able to track the continuously varying impulse response in this experiment. The estimation process involves predicting the impulse response over short, overlapping windows (an example is shown in Figure~\ref{fig:ltv_experiment1_single_window_fit}) and then stitching these estimates together to form the complete time-varying response. As shown in Figure~\ref{fig:ltv_experiment1_stitched_fit}, the posterior mean of the final stitched impulse response tracks the ground truth reasonably closely.

This experiment highlights the power of the GP as a regularization tool. By imposing a prior belief that the impulse response should vary smoothly over time, the ill-posed regression problem becomes solvable. The model learns to leverage the temporal structure of the signals to infer how the system evolves, effectively "filling in the gaps" where direct observation is impossible. The successful application, even when the experimental noise model diverges from the theoretical assumptions, underscores the flexibility and practical utility of this Bayesian approach for analyzing complex, non-stationary systems.

As in the single-observation LTI experiment, this result should therefore be interpreted primarily as a demonstration of robustness under misspecification. Here the posterior spread reflects a combination of inferential uncertainty, regularization, and mismatch compensation, rather than a literal validation of intrinsic physical randomness in the channel.

\section{Conclusion}

This paper introduced a unified Bayesian framework for modeling and estimating linear time-varying systems. By representing uncertainty directly at the level of the latent impulse response, the framework provides a posterior interpretation that connects deterministic LTI systems, fully stochastic channels, and LTIE laws over impulse responses. A particular strength of this viewpoint is that intrinsic channel variability and epistemic uncertainty are unified through the same posterior representation, while additive observation noise can be treated as a separate observation layer when needed. When the true data-generating mechanism instead includes a separate additive observation-noise term, the resulting posterior should be interpreted as an uncertainty-aware inferential representation rather than as a literal claim about the physical source of all randomness.

We demonstrated the practical utility and robustness of this framework through a series of experiments. Our results showed that by leveraging modern tools like Bayesian neural networks and variational inference, we can solve the classic, ill-posed deconvolution problem for an LTI system from a single noisy input-output pair, successfully quantifying the estimation uncertainty. In a simulated geophysical application, our Bayesian approach showed a lower overall error floor than the classical CCF stacking baseline. Furthermore, we demonstrated that by incorporating a Gaussian Process as a structured temporal prior, our framework can track a continuously varying LTV impulse response in the simulated setting considered here, a task that is challenging for many standard techniques.

The primary implication of this work is that the fusion of classical system theory with probabilistic machine learning offers a more complete and honest assessment of system behavior. The ability to propagate uncertainty from the impulse response to any derived quantity such as a denoised signal or a cross-correlation function is critical for robust decision-making in fields like control theory, communications, and geophysics.

Despite these promising results, we acknowledge certain limitations that pave the way for future research. The variational inference methods employed are more computationally intensive than their classical counterparts, involving iterative optimization and sampling, whereas methods like CCF stacking are non-iterative. The quality of the LTV regression is also contingent on the choice of a suitable prior. Future work should focus on three key areas. First, exploring more advanced posterior approximations, such as normalizing flows, could capture complex, non-Gaussian uncertainties. Second, applying the framework to real-world datasets from seismology or wireless communications will be essential for validation. Finally, extending this probabilistic approach to more complex system classes, such as Multiple-Input Multiple-Output (MIMO) or specific nonlinear systems, presents an exciting avenue for continued research. More broadly, the experiments with additive observation noise show that the framework remains practically useful even under deliberate model misspecification.

\bibliographystyle{elsarticle-harv}

\bibliography{refs}

@inproceedings{tensorflow2015-whitepaper,
  author    = {Mart{\'i}n Abadi and Paul Barham and Jianmin Chen and Zhifeng Chen and
               Andy Davis and Jeffrey Dean and Matthieu Devin and Sanjay Ghemawat and
               Geoffrey Irving and Michael Isard and Manjunath Kudlur and Josh Levenberg and
               Rajat Monga and Sherry Moore and Derek G. Murray and Benoit Steiner and
               Paul Tucker and Vijay Vasudevan and Pete Warden and Martin Wicke and
               Yuan Yu and Xiaoqiang Zheng},
  title     = {TensorFlow: A System for Large-Scale Machine Learning},
  booktitle = {12th USENIX Symposium on Operating Systems Design and Implementation (OSDI 16)},
  pages     = {265--283},
  year      = {2016},
  url       = {https://www.usenix.org/system/files/conference/osdi16/osdi16-abadi.pdf}
}

@article{Dillon2017TFD,
  author        = {Joshua V. Dillon and Ian Langmore and Dustin Tran and Eugene Brevdo and Srinivas Vasudevan and Dave Moore and Brian Patton and Alex Alemi and Matt Hoffman and Rif A. Saurous},
  title         = {TensorFlow Distributions},
  journal       = {arXiv preprint arXiv:1711.10604},
  year          = {2017},
  eprint        = {1711.10604},
  archivePrefix = {arXiv},
  primaryClass  = {stat.ML},
  url           = {https://arxiv.org/abs/1711.10604}
}

@article{Kingma2013,
  title={Auto-Encoding Variational Bayes},
  author={Diederik P. Kingma and Max Welling},
  journal={CoRR},
  year={2013},
  volume={abs/1312.6114},
  url={https://arxiv.org/abs/1312.6114}
}

@inproceedings{Loshchilov2017,
  author    = {Ilya Loshchilov and Frank Hutter},
  title     = {SGDR: Stochastic Gradient Descent with Warm Restarts},
  booktitle = {International Conference on Learning Representations},
  year      = {2017},
  url       = {https://openreview.net/forum?id=Skq89Scxx}
}

@article{Bello1963,
  author  = {P. A. Bello},
  title   = {Characterization of Randomly Time-Variant Linear Channels},
  journal = {IRE Transactions on Communications Systems},
  year    = {1963},
  volume  = {11},
  number  = {4},
  pages   = {360--393},
  month   = {Dec},
  doi     = {10.1109/TCOM.1963.1088793}
}

@inproceedings{Aubel2015,
  author    = {Celine Aubel and Helmut B{\"o}lcskei},
  title     = {Density Criteria for the Identification of Linear Time-Varying Systems},
  booktitle = {2015 IEEE International Symposium on Information Theory (ISIT)},
  year      = {2015},
  month     = {June},
  pages     = {2568--2572},
  doi       = {10.1109/ISIT.2015.7282920}
}

@inbook{Kozachek2024,
  author    = {Olga Kozachek and Nikolay Nikolaev and Olga Slita and Alexey Bobtsov},
  title     = {Parameter Identification Algorithm for a LTV System with Partially Unknown State Matrix},
  booktitle = {Mathematical Modelling, Computational Intelligence Techniques and Renewable Energy},
  year      = {2024},
  month     = {Sep},
  pages     = {306--319},
  isbn      = {978-3-031-71359-0},
  doi       = {10.1007/978-3-031-71360-6_23}
}

@article{Lin2020,
  author  = {Sen Lin and Hang Wang and Junshan Zhang},
  title   = {System Identification via Meta-Learning in Linear Time-Varying Environments},
  year    = {2020},
  month   = {Oct},
  journal = {arXiv preprint arXiv:2010.14664},
  archivePrefix = {arXiv},
  eprint  = {2010.14664},
  doi     = {10.48550/arXiv.2010.14664}
}

@article{Kailath1962,
  author    = {T. Kailath},
  title     = {Measurements on Time-Variant Communication Channels},
  journal   = {IRE Transactions on Information Theory},
  volume    = {8},
  number    = {5},
  pages     = {229--236},
  year      = {1962},
  doi       = {10.1109/TIT.1962.1057748}
}

@article{Yoo2005,
  author    = {Do-Sik Yoo and W. E. Stark},
  title     = {Characterization of WSSUS Channels: Normalized Mean Square Covariance},
  journal   = {IEEE Transactions on Wireless Communications},
  volume    = {4},
  number    = {4},
  pages     = {1575--1584},
  year      = {2005},
  doi       = {10.1109/TWC.2004.843046}
}

@article{Priestley1965,
  author  = {M. B. Priestley},
  title   = {Evolutionary Spectra and Non-Stationary Processes},
  journal = {Journal of the Royal Statistical Society: Series B (Methodological)},
  volume  = {27},
  number  = {2},
  pages   = {204--237},
  year    = {1965},
  doi     = {10.1111/j.2517-6161.1965.tb01488.x},
  url     = {https://doi.org/10.1111/j.2517-6161.1965.tb01488.x}
}

@article{Dahlhaus1997,
  author  = {Rainer Dahlhaus},
  title   = {Fitting Time Series Models to Nonstationary Processes},
  journal = {The Annals of Statistics},
  volume  = {25},
  number  = {1},
  pages   = {1--37},
  year    = {1997},
  doi     = {10.1214/AOS/1034276620},
  url     = {https://doi.org/10.1214/AOS/1034276620}
}

@article{Pillonetto2023,
  author  = {G. Pillonetto and Lennart Ljung},
  title   = {Full Bayesian Identification of Linear Dynamic Systems Using Stable Kernels},
  journal = {Proceedings of the National Academy of Sciences of the United States of America},
  volume  = {120},
  pages   = {e2218197120},
  year    = {2023},
  doi     = {10.1073/pnas.2218197120},
  url     = {https://www.pnas.org/doi/10.1073/pnas.2218197120}
}

@phdthesis{Darwish2017,
  author    = {M.A.H. Darwish},
  title     = {Bayesian Identification of Linear Dynamic Systems: Synthesis of Kernels in the LTI Case and Beyond},
  school    = {Technische Universiteit Eindhoven},
  year      = {2017},
  month     = {Oct},
  day       = {10},
  address   = {Eindhoven, The Netherlands},
  isbn      = {978-90-386-4352-6},
  note      = {Proefschrift},
  url       = {https://pure.tue.nl/ws/files/77531824/20171010_Darwish.pdf}
}

@article{Narendra1990,
  author    = {K. S. Narendra and K. Parthasarathy},
  title     = {Identification and Control of Dynamical Systems Using Neural Networks},
  journal   = {IEEE Transactions on Neural Networks},
  volume    = {1},
  number    = {1},
  pages     = {4--27},
  year      = {1990},
  doi       = {10.1109/72.80202}
}

@book{Nelles2020,
  author    = {Oliver Nelles},
  title     = {Nonlinear System Identification: From Classical Approaches to Neural Networks, Fuzzy Models, and Gaussian Processes},
  edition   = {2nd},
  publisher = {Springer},
  year      = {2020},
  doi       = {10.1007/978-3-030-47439-3},
  isbn      = {978-3-030-47438-6}
}

@article{Jospin2022,
  author    = {Laurent Valentin Jospin and Hamid Laga and Farid Boussaid and Wray Buntine and Mohammed Bennamoun},
  title     = {Hands-On Bayesian Neural Networks—A Tutorial for Deep Learning Users},
  journal   = {IEEE Computational Intelligence Magazine},
  volume    = {17},
  number    = {2},
  pages     = {29--48},
  year      = {2022},
  doi       = {10.1109/MCI.2022.3155327}
}

@article{Zhou2022,
  author  = {Hongpeng Zhou and Chahine Ibrahim and Wei Xing Zheng and Wei Pan},
  title   = {Sparse Bayesian Deep Learning for Dynamic System Identification},
  journal = {Automatica},
  volume  = {144},
  pages   = {110489},
  year    = {2022},
  doi     = {10.1016/j.automatica.2022.110489},
  url     = {https://doi.org/10.1016/j.automatica.2022.110489}
}

@article{Stock2024,
  author    = {Simon Stock and Davood Babazadeh and Christian Becker and Spyros Chatzivasileiadis},
  title     = {Bayesian Physics-informed Neural Networks for System Identification of Inverter-Dominated Power Systems},
  journal   = {Electric Power Systems Research},
  volume    = {235},
  pages     = {110860},
  year      = {2024},
  issn      = {0378-7796},
  doi       = {10.1016/j.epsr.2024.110860},
  url       = {https://www.sciencedirect.com/science/article/pii/S0378779624007466}
}

@inproceedings{Damianou2013,
  title     = {Deep {G}aussian Processes},
  author    = {Andreas Damianou and Neil D. Lawrence},
  booktitle = {Proceedings of the Sixteenth International Conference on Artificial Intelligence and Statistics},
  pages     = {207--215},
  year      = {2013},
  editor    = {Carlos M. Carvalho and Pradeep Ravikumar},
  volume    = {31},
  series    = {Proceedings of Machine Learning Research},
  address   = {Scottsdale, Arizona, USA},
  month     = {29 Apr--01 May},
  publisher = {PMLR},
  url       = {https://proceedings.mlr.press/v31/damianou13a.html}
}

@inproceedings{Sendera2025,
  author    = {Marcin Sendera and Amin Sorkhei and Tomasz Ku{\'s}mierczyk},
  title     = {Revisiting the Equivalence of Bayesian Neural Networks and Gaussian Processes: On the Importance of Learning Activations},
  booktitle = {Proceedings of the Forty-first Conference on Uncertainty in Artificial Intelligence},
  pages     = {3675--3700},
  year      = {2025},
  editor    = {Silvia Chiappa and Sara Magliacane},
  volume    = {286},
  series    = {Proceedings of Machine Learning Research},
  publisher = {PMLR},
  url       = {https://proceedings.mlr.press/v286/sendera25a.html}
}

@article{Bhouri2022,
  author  = {Mohamed Aziz Bhouri and Paris Perdikaris},
  title   = {Gaussian Processes Meet NeuralODEs: A Bayesian Framework for Learning the Dynamics of Partially Observed Systems from Scarce and Noisy Data},
  journal = {Philosophical Transactions of the Royal Society A: Mathematical, Physical and Engineering Sciences},
  volume  = {380},
  number  = {2229},
  pages   = {20210201},
  year    = {2022},
  month   = jun,
  doi     = {10.1098/rsta.2021.0201},
  url     = {https://doi.org/10.1098/rsta.2021.0201}
}

@article{Hallemans2020,
  author  = {No{\"e}l Hallemans and John Lataire and Rik Pintelon},
  title   = {Nonparametric Identification of Linear Time-Varying Systems using Gaussian Process Regression},
  journal = {IFAC-PapersOnLine},
  volume  = {53},
  number  = {2},
  pages   = {1001--1006},
  year    = {2020},
  doi     = {10.1016/j.ifacol.2020.12.1277},
  url     = {https://doi.org/10.1016/j.ifacol.2020.12.1277}
}

@article{Susik2023,
  author    = {M. Susik and I. F. Sbalzarini},
  title     = {Variational Inference Accelerates Accurate DNA Mixture Deconvolution},
  journal   = {Forensic Science International: Genetics},
  volume    = {65},
  pages     = {102890},
  year      = {2023},
  month     = {Jul},
  doi       = {10.1016/j.fsigen.2023.102890},
  note      = {Epub 2023 May 20},
  issn      = {1878-0326},
  pmid      = {37257308}
}

@inproceedings{Zhang2013,
  author    = {Ganchi Zhang and Nick Kingsbury},
  title     = {Fast L0-based Image Deconvolution with Variational Bayesian Inference and Majorization-Minimization},
  booktitle = {Proceedings of the 2013 IEEE Global Conference on Signal and Information Processing (GlobalSIP)},
  pages     = {1081--1084},
  year      = {2013},
  publisher = {IEEE},
  doi       = {10.1109/GlobalSIP.2013.6737081},
  url       = {https://www-sigproc.eng.cam.ac.uk/foswiki/pub/Main/GZ243/06737081.pdf}
}

@inproceedings{Wen2018,
  author    = {Yeming Wen and Paul Vicol and Jimmy Ba and Dustin Tran and Roger Grosse},
  title     = {Flipout: Efficient Pseudo-Independent Weight Perturbations on Mini-Batches},
  booktitle = {International Conference on Learning Representations},
  year      = {2018},
  url       = {https://openreview.net/forum?id=rJNpifWAb}
}

@inproceedings{Rezende2014,
  title     = {Stochastic Backpropagation and Approximate Inference in Deep Generative Models},
  author    = {Danilo Jimenez Rezende and Shakir Mohamed and Daan Wierstra},
  booktitle = {Proceedings of the 31st International Conference on Machine Learning},
  pages     = {1278--1286},
  year      = {2014},
  editor    = {Eric P. Xing and Tony Jebara},
  volume    = {32},
  series    = {Proceedings of Machine Learning Research},
  address   = {Beijing, China},
  month     = {22--24 Jun},
  publisher = {PMLR},
  url       = {http://proceedings.mlr.press/v32/rezende14.html}
}

@article{EarthquakesUCR2018,
  author    = {Hoang Anh Dau and Anthony Bagnall and Kaveh Kamgar and
               Chin{-}Chia Michael Yeh and Yan Zhu and Shaghayegh Gharghabi and
               Chotirat Ann Ratanamahatana and Eamonn Keogh},
  title     = {The {UCR} Time Series Archive},
  journal   = {IEEE/CAA Journal of Automatica Sinica},
  volume    = {6},
  number    = {6},
  pages     = {1293--1305},
  year      = {2019},
  doi       = {10.1109/JAS.2019.1911747},
  url       = {https://doi.org/10.1109/JAS.2019.1911747},
  note      = {Used for the Earthquakes dataset from the UCR Archive.}
}

@article{Bensen2007,
  author  = {G. D. Bensen and M. H. Ritzwoller and M. P. Barmin and A. L. Levshin and F. Lin and M. P. Moschetti and N. M. Shapiro and Y. Yang},
  title   = {Processing seismic ambient noise data to obtain reliable broad-band surface wave dispersion measurements},
  journal = {Geophysical Journal International},
  volume  = {169},
  number  = {3},
  pages   = {1239--1260},
  year    = {2007},
  month   = jun,
  doi     = {10.1111/j.1365-246X.2007.03374.x},
  url     = {https://academic.oup.com/gji/article/169/3/1239/626431}
}

@article{Ekstrom2009,
  author  = {G{\"o}ran Ekstr{\"o}m and Geoffrey A. Abers and Spahr C. Webb},
  title   = {Determination of surface-wave phase velocities across {USArray} from noise and {Aki's} spectral formulation},
  journal = {Geophysical Research Letters},
  volume  = {36},
  pages   = {L18301},
  year    = {2009},
  month   = sep,
  doi     = {10.1029/2009GL039131},
  url     = {https://doi.org/10.1029/2009GL039131}
}

@article{Snieder2004,
  author  = {Roel Snieder},
  title   = {Extracting the Green's Function from the Correlation of Coda Waves: A Derivation Based on Stationary Phase},
  journal = {Physical Review E},
  volume  = {69},
  number  = {4},
  pages   = {046610},
  year    = {2004},
  doi     = {10.1103/PhysRevE.69.046610},
  url     = {https://doi.org/10.1103/PhysRevE.69.046610}
}

@article{SanchezSesma2006,
  author  = {F. J. S{\'a}nchez-Sesma and J. A. P{\'e}rez-Ruiz and M. Campillo and F. Luz{\'o}n},
  title   = {Retrieval of the Green's Function from Cross Correlation: The Canonical Elastic Problem},
  journal = {Bulletin of the Seismological Society of America},
  volume  = {96},
  number  = {3},
  pages   = {1182--1191},
  year    = {2006},
  doi     = {10.1785/0120050181},
  url     = {https://doi.org/10.1785/0120050181}
}

@article{Aki1957,
  author  = {Keiiti Aki},
  title   = {Space and Time Spectra of Stationary Stochastic Waves, with Special Reference to Microtremors},
  journal = {Bulletin of the Earthquake Research Institute, University of Tokyo},
  volume  = {35},
  number  = {3},
  pages   = {415--456},
  year    = {1957},
  doi     = {10.15083/0000033938},
  url     = {https://doi.org/10.15083/0000033938}
}

@article{Willems2005PE,
  author  = {Jan C. Willems and Paolo Rapisarda and Ivan Markovsky and Bart L. M. De Moor},
  title   = {A Note on Persistency of Excitation},
  journal = {Systems \& Control Letters},
  volume  = {54},
  number  = {4},
  pages   = {325--329},
  year    = {2005},
  doi     = {10.1016/j.sysconle.2004.09.003},
  url     = {https://eprints.soton.ac.uk/262195/1/PersistencyExcitation.pdf}
}

@incollection{BerteroBoccacci2005,
  author    = {Mario Bertero and Patrizia Boccacci},
  title     = {Image Deconvolution},
  booktitle = {Microscopy Techniques},
  publisher = {Springer},
  year      = {2005},
  pages     = {355--378},
  doi       = {10.1007/1-4020-3616-7_17},
  url       = {https://person.dibris.unige.it/bertero-mario/papers-PDF/ASI.pdf}
}

@misc{TFPConvolution1DFlipoutDocs,
  author       = {{TensorFlow Authors}},
  title        = {tfp.layers.Convolution1DFlipout},
  year         = {2023},
  howpublished = {\url{https://www.tensorflow.org/probability/api_docs/python/tfp/layers/Convolution1DFlipout}},
  note         = {TensorFlow Probability API documentation, accessed 2026-03-21}
}

@book{SimonAlouini2005,
  author    = {Marvin K. Simon and Mohamed-Slim Alouini},
  title     = {Digital Communication over Fading Channels},
  edition   = {2nd},
  publisher = {Wiley},
  year      = {2005},
  isbn      = {978-0-471-64953-3}
}

\clearpage

\appendix
\section{Experimental Settings}
\label{app:experimental_settings}

For reproducibility, we summarize here the main settings used to generate the reported experiments and uncertainty visualizations; the full code, notebooks, and environment details are available at \url{https://github.com/yaniv-shulman/bayes-ltv}.

\paragraph{\textbf{Single-observation LTIE experiment}}
The input signal was a single series from the UCR ``Earthquakes'' dataset. The ground-truth FIR had length $16$ and was generated with \texttt{firwin} using cutoff bands $[0.25, 0.40, 0.50, 0.95]$, width $0.05$, and \texttt{pass\_zero=False}. The observed output was obtained by valid convolution followed by additive Gaussian noise with standard deviation $0.5$. The approximate posterior was a mean-field Gaussian over the convolution kernel, with prior scale parameter $\alpha=1$ (equivalently, prior standard deviation $1/\sqrt{\text{kernel\_size}}$), trained for $1500$ epochs with Adam and a cosine warmup/decay schedule, warming the learning rate from $0$ to $0.5$ over $4\%$ of the epochs. For the single-observation setting, the lone observation pair was repeated to a computational batch size of $1024$ for Monte Carlo estimation. The FIR uncertainty bands use $\pm 2$ posterior standard deviations, and the transfer-function and CCF summaries in the appendix are based on $500$ posterior samples.

A repeated synthetic calibration study was also run with $100$ outer replications, known noise standard deviation $0.5$, and pointwise tapwise coverage checks for the approximate variational posterior. For $1$ independent observation pair, nominal $90\%$ and $95\%$ intervals achieved empirical coverages of $97.5\%$ and $99.4\%$, with mean interval widths $0.124$ and $0.147$. For $20$ independent pairs, both nominal levels achieved $100\%$ empirical coverage with mean widths $0.090$ and $0.107$, and for $50$ independent pairs the corresponding mean widths were $0.089$ and $0.106$. Thus, in this synthetic LTIE setting, the pointwise intervals were conservative, while uncertainty contracted substantially as more independent observations were added.

\paragraph{\textbf{ANT experiment}}
The synthetic ANT dataset comprised up to $15{,}000$ IID receiver pairs, each of length $1200$ samples, with sample rate $20$ Hz, receiver spacing $9000$ m, and additive Gaussian noise standard deviation $0.1$. Each pair was generated from $10{,}000$ decaying sinusoidal source pulses of length $50$ samples, with random amplitude in $[-1.5,1.5]$, random frequency in $[0.1,10]$ Hz, and decay constant $\tau \in [0.5,2.5]$; the target dispersive curve was anchored at $(0.1~\text{Hz},3500~\text{m/s})$, $(1~\text{Hz},3000~\text{m/s})$, and $(10~\text{Hz},1000~\text{m/s})$. The Bayesian LTIE estimator used a batch size of 2000 at minimum, $\alpha=1$, $500$ epochs, and Adam with the same $4\%$ warmup/cosine schedule and target learning rate $0.1$. In the full-precision experiment, the Gaussian data-fit weight used a residual-estimated observation-noise scale. Performance was evaluated for $1000,2000,\ldots,15000$ pairs, on a velocity grid from $500$ to $5000$ m/s with $226$ points and a frequency grid from $0.1$ to $3.0$ Hz with $300$ points. The standard experiment in the main text did not use 1-bit quantization, and the Bayesian MIR was trained directly on non-whitened signal pairs. For the 1-bit experiment, the same settings were used except that the batch size was increased to $4000$. The velocity-misfit uncertainty bands are pointwise $2.5\%$--$97.5\%$ sample percentiles.

\paragraph{\textbf{LTV experiment}}
The ground-truth time-varying FIR used $16$ taps and was formed by smoothly interpolating between three FIRs generated with \texttt{firwin} cutoffs $[0.05,0.4,0.6,0.95]$, $[0.2,0.93]$, and $[0.09,0.89]$, each with width $0.05$ and \texttt{pass\_zero=False}. The observed signal used additive Gaussian noise with standard deviation $0.2$. Local inference windows had length $32$ and were extracted at unit stride; the $291$ unique windows were tiled twice for training, yielding tensors of shapes $(582,32,1)$, $(582,32,16)$, and $(582,32)$. The amortized variational model used $64$ initial CNN filters and a GP prior with RBF amplitude $2.0$, length scale $16$, kernel jitter $10^{-5}$, and KL scaling $5 \times 10^{-5}$. Training used Adam with cosine warmup/decay from learning rate $0$ to $0.003$, warmup over $4\%$ of $7000$ epochs, and full-batch optimization. The final global estimate was formed by averaging overlapping local impulse-response predictions during stitching.

\section{Impulse Response Regression from a Single Observation}

\clearpage

\section{Application to Simulated Ambient Noise Tomography}

\begin{figure}[!ht]
    \centering
    \includegraphics[width=\linewidth]{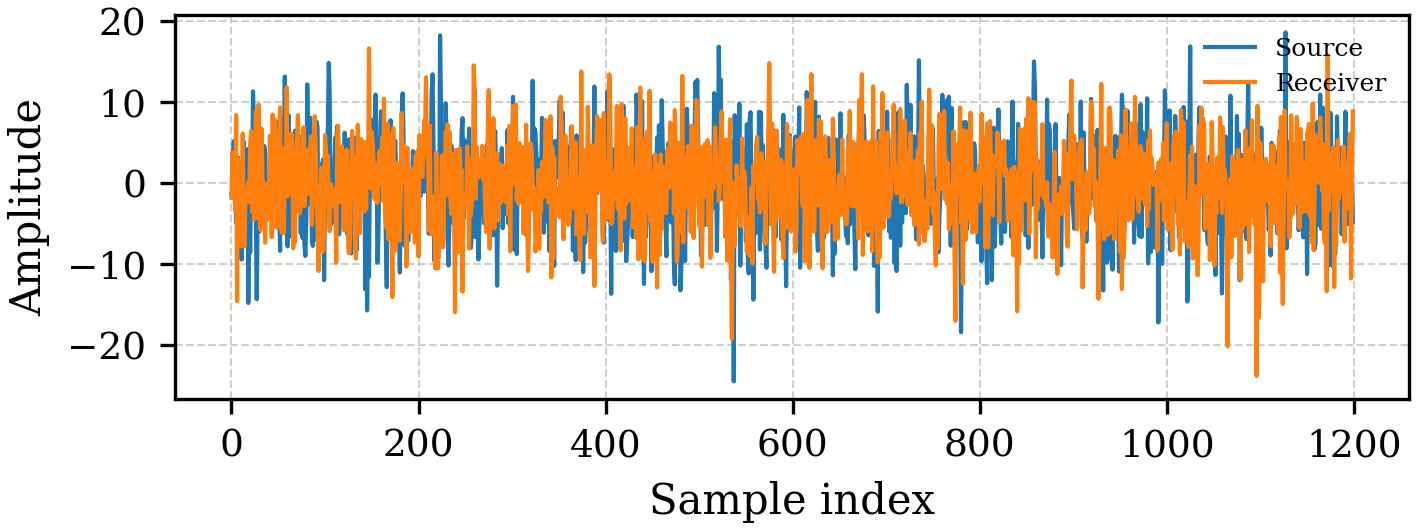}
    \caption{An example observation pair from the synthetic ANT simulation. Each signal represents the noisy superposition of waves from multiple sources, as recorded at two different receivers.}
    \label{fig:ant_example_pair}
\end{figure}

\begin{figure}[!h]
    \centering
    \begin{subfigure}[b]{.47\linewidth}
        \includegraphics[width=\linewidth]{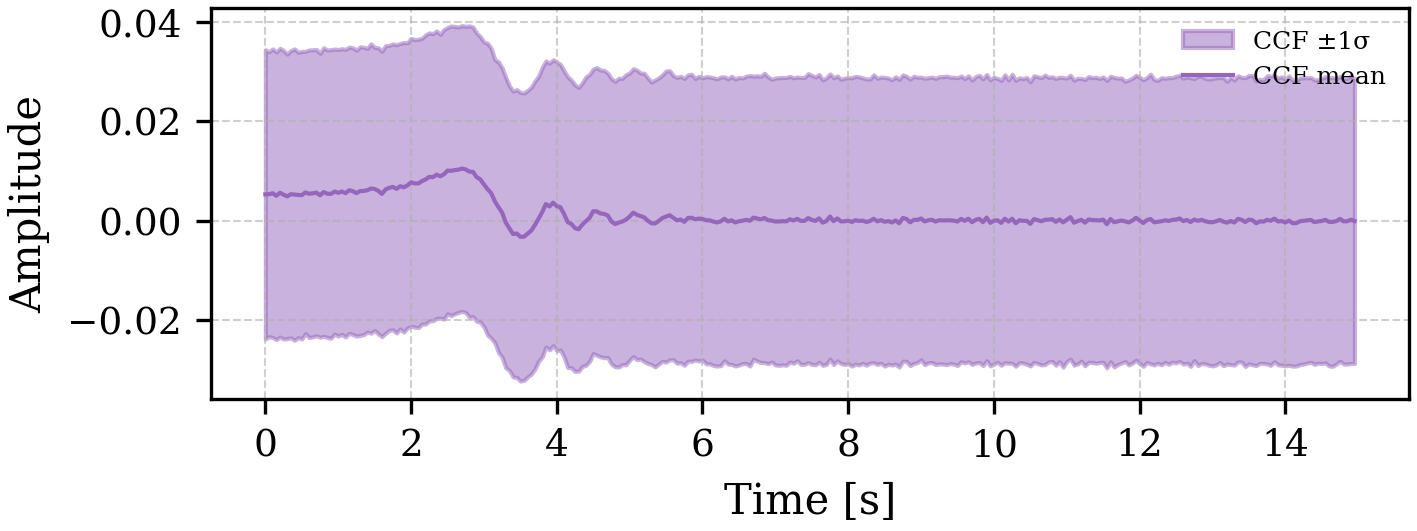}
        \subcaption{The mean of the estimated Green's function computed by classical CCF stacking. The shaded region shows the empirical mean $\pm 1$ empirical standard deviation across the estimates.}
        \label{fig:ant_ccf_estimate}
    \end{subfigure}
    \hfill
    \begin{subfigure}[b]{.47\linewidth}
        \includegraphics[width=\linewidth]{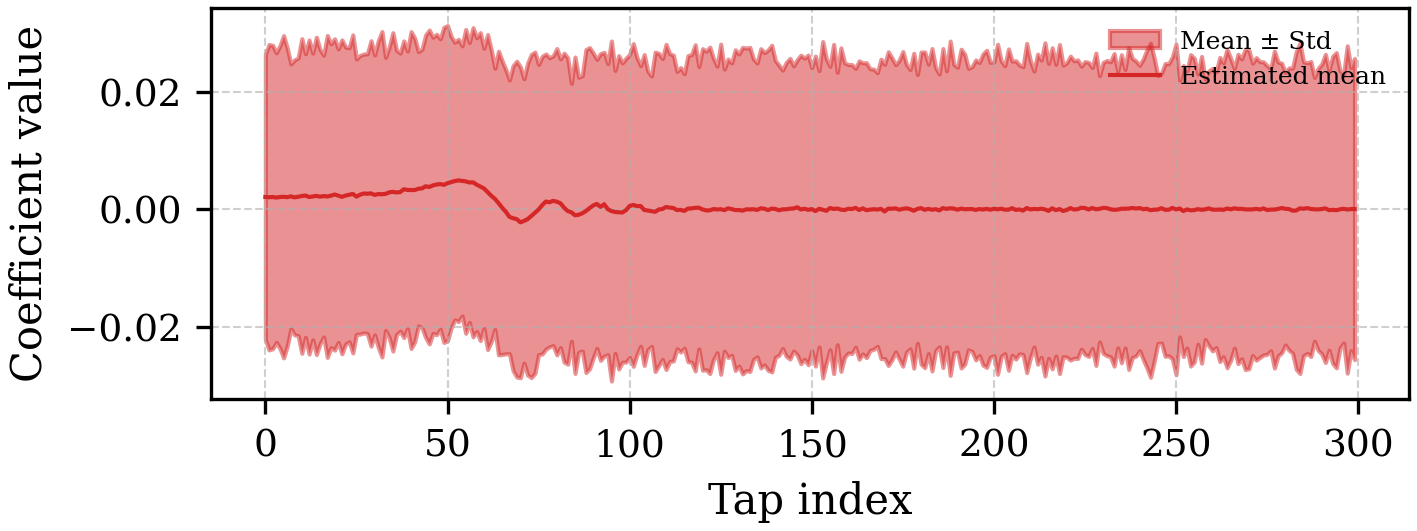}
        \subcaption{The posterior mean of the impulse response computed by our Bayesian model. The shaded region shows the posterior mean $\pm 1$ posterior standard deviation.}
        \label{fig:ant_mir_estimate}
    \end{subfigure}
    \caption{Visualization of the estimated Green's function from (a) classical CCF stacking and (b) our Bayesian model, using 12000 signal pairs in a simulated ANT experiment.}
\end{figure}

\clearpage

\section{Regression of a Non-Stationary Impulse Response}

\end{document}